\documentclass[vecphys]{svmult}

\usepackage{graphicx}        
\usepackage{multicol}        
\usepackage[bottom]{footmisc}

\def\msun{\mbox{M$_{\odot}$}}
\def\msunh{\mbox{M$_{\odot}$h$^{-1}$}}
\def\Lcdm{\mbox{$\Lambda$CDM}}
\def\mpch{\mbox{Mpch$^{-1}$}}
\def\ae{\mbox{$a_{eq}$}}
\def\te{\mbox{$t_{eq}$}}
\def\tcr{\mbox{$t_{cross}$}}
\def\ac{\mbox{$a_{cross}$}}
\def\trec{\mbox{$t_{rec}$}}
\def\arec{\mbox{$a_{rec}$}}

\def\lp{\mbox{$\lambda_{p}$}}
\def\tp{\mbox{$t_{press}$}}
\def\tg{\mbox{$t_{grav}$}}
\def\lh{\mbox{$L_{H}$}}
\def\deltaX{\mbox{$\delta(\bf{x})$}}
\def\dk{\mbox{$\delta_k$}}
\def\sms{\mbox{$\sigma_M^2$}}
\def\sm{\mbox{$\sigma_M$}}

\def\mnras{\mbox{MNRAS}}
\def\apj{\mbox{ApJ}}
\def\aap{\mbox{A\&A}}
\def\aj{\mbox{AJ}}
\def\apj{\mbox{ApJ}}
\def\araa{\mbox{ARA\&A}}

\def\grtsim{{_ >\atop{^\sim}}}
\def\lesssim{{_ <\atop{^\sim}}}

\begin{document}

\title*{Understanding Galaxy Formation and Evolution}
\author{Vladimir Avila-Reese\inst{1}}
\institute{Instituto de Astronom\'{\i}a, Universidad Nacional Aut\'onoma
de M\'exico, A.P. 70-264, 04510, M\'exico,D.F. 
\texttt{avila@astroscu.unam.mx}}
%
%
\maketitle

The old dream of integrating into one the study of micro and macrocosmos is now 
a reality. Cosmology, astrophysics, and particle physics intersect in a 
scenario (but still not a theory) of cosmic structure formation and evolution 
called $\Lambda$ Cold Dark Matter (\Lcdm) model. This scenario emerged mainly to 
explain the origin of galaxies. In these lecture notes, I first present a review of the 
main galaxy properties, highlighting the questions that any theory of galaxy 
formation should explain. Then, the cosmological framework and the
main aspects of primordial perturbation generation and evolution are
pedagogically detached. Next, I focus on the ``dark side'' of galaxy 
formation, presenting a review on \Lcdm\ halo assembling and properties,
and on the main candidates for non--baryonic dark matter. It is shown
how the nature of elemental particles can influence on the features
of galaxies and their systems.
Finally, the complex processes of baryon dissipation inside the non--linearly 
evolving CDM halos, formation of disks and spheroids, and transformation of gas
into stars are briefly described, remarking on the possibility of a few
driving factors and parameters able to explain the main body of galaxy
properties. A summary and a discussion of some of the issues and open problems
of the \Lcdm\ paradigm are given in the final part of these notes. 

\section{Introduction}
\label{sec:1}

Our vision of the cosmic world and in particular of the whole Universe has been 
changing dramatically in the last century. As we will see, galaxies were repeatedly 
the main protagonist in the scene of these changes. It is about 80 years since 
E. Hubble established the nature of galaxies as gigantic self-bound stellar systems
and used their kinematics to show that the Universe as a whole is 
expanding uniformly at the present time. Galaxies, as the building blocks 
of the Universe, are also tracers
of its large--scale structure and of its evolution in the last 13 Gyrs or more. 
By looking inside galaxies we find that they are the arena where stars form, 
evolve and collapse
in constant interaction with the interstellar medium (ISM), a complex mix of gas
and plasma, dust, radiation, cosmic rays, and magnetics fields. The center
of a significant fraction of galaxies harbor supermassive black holes.  When
these ``monsters'' are fed with infalling material, the accretion disks
around them release, mainly through powerful plasma jets, the largest amounts of 
energy known in astronomical objects. This phenomenon of Active Galactic Nuclei
(AGN) was much more frequent in the past than in the present, being the high--redshift 
quasars (QSO's) the most powerful incarnation of the AGN phenomenon. But the
most astonishing surprise of galaxies comes from the fact that luminous matter 
(stars, gas, AGN's, etc.) is only a tiny fraction ($\sim 1-5\%$) of all the mass
measured in galaxies and the giant halos around them. What this 
dark component of galaxies is made of? This is one of the most acute enigmas of modern
science.

Thus, exploring and understanding galaxies is of paramount interest to
cosmology, high--energy and particle physics, gravitation theories, and,
of course, astronomy and astrophysics. As astronomical objects, among
other questions, we would like to know how do they take shape and evolve, 
what is the origin of their diversity and scaling laws, why they cluster 
in space as observed, following a sponge--like structure, what 
is the dark component that predominates in their masses. By answering to these
questions we would able also to use galaxies as a true link between the 
observed universe and the properties of the early universe, and as 
physical laboratories for testing fundamental theories. 

The content of these notes is as follows. In \S 2 a review on main
galaxy properties and correlations is given. By following an analogy
with biology, the taxonomical, anatomical, ecological and genetical
study of galaxies is presented. The observational inference of dark 
matter existence, and the baryon budget in galaxies and in the
Universe is highlighted. Section 3 is dedicated to a pedagogical 
presentation of the basis of cosmic structure formation theory
in the context of the $\Lambda$ Cold Dark Matter (\Lcdm) paradigm.
The main questions to be answered are: why CDM is invoked to explain 
the formation of galaxies? How is explained the origin of the seeds 
of present--day cosmic structures? How these seeds evolve?.
In \S 4 an updated review of the main results on properties and 
evolution of CDM halos is given, with emphasis on the aspects that
influence the propertied of the galaxies expected to be formed
inside the halos. A short discussion on dark matter candidates
is also presented (\S\S 4.2).  The main ingredients of disk
and spheroid galaxy formation are reviewed and discussed in 
\S 5.  An attempt to highlight the main drivers of the Hubble
and color sequences of galaxies is given in \S\S 5.3. Finally,
some selected issues and open problems in the field are resumed
and discussed in \S 6.

\section{Galaxy properties and correlations}
\label{sec:2}

During several decades galaxies were considered basically as self--gravitating 
stellar systems so that the study of their physics was a domain of Galactic 
Dynamics. Galaxies in the local Universe are indeed mainly conglomerates of 
hundreds of millions to trillions of stars {\it supported against 
gravity either by rotation or by random motions}. In the former case, the system
has the shape of a {\it flattened disk}, where most of the material is on 
circular orbits at radii that are the minimal ones allowed by the 
specific angular momentum of the material. Besides, disks are dynamically fragile 
systems, unstable to perturbations. Thus, the mass distribution along the disks 
is the result of the specific angular momentum distribution of the material from 
which the disks form, and of the posterior dynamical (internal and external) 
processes. In the latter case, the shape of the galactic system is a concentrated 
{\it spheroid/ellipsoid}, with mostly (disordered) radial orbits. The spheroid
is dynamically hot, stable to perturbations. Are the properties of the stellar
populations in the disk and spheroid systems different?


\paragraph{Stellar populations}

Already in the 40's, W. Baade discovered that according to the ages, metallicities, 
kinematics and spatial distribution of the stars in our Galaxy, they separate in 
two groups: 1) Population I stars, which populate the plane of the disk; 
their ages do not go beyond 10 Gyr --a fraction of them in fact are young 
($\lesssim 10^6$ yr) luminous O,B stars mostly in the spiral arms, and their 
metallicites are close to the solar one, $Z\approx 2\%$;  2) Population II stars, 
which are located in the spheroidal component of the 
Galaxy (stellar halo and partially in the bulge), 
where velocity dispersion (random motion) is higher than rotation velocity (ordered 
motion); they are old stars ($>10$ Gyr) with very low metallicities, on the average 
lower by two orders of magnitude than Population I stars. In between Pop's I and II 
there are several stellar subsystems.
\footnote{Astronomers suspect also the existence of non--observable
Population III of pristine stars with zero metallicities, formed
in the first molecular clouds $\sim 4 \ 10^{8}$ yrs ($z\sim 20$) after the Big Bang. 
These stars are thought to be very massive, so that in scaletimes of 
1Myr they exploded, injected a big amount of energy to the primordial gas and started
to reionize it through expanding cosmological HII regions (see e.g., \cite{BL04,CF05}
for recent reviews on the subject).}.

Stellar populations are true fossils of the galaxy assembling process. The 
differences between them evidence differences in the formation and evolution of 
the galaxy components. The Pop II stars, being old, of low metallicity, and
dominated by random motions (dynamically hot), had to form early in the 
assembling history of galaxies and through violent processes. In the meantime,
the large range of ages of Pop I stars, but on average younger than the Pop
II stars, indicates a slow star formation process that 
continues even today in the disk plane. Thus, the common wisdom says that
{\it spheroids form early in a violent collapse (monolithic or major
merger), while disks assemble by continuous infall of gas rich in angular 
momentum, keeping a self--regulated SF process.}


\paragraph{Interstellar Medium (ISM)}

Galaxies are not only conglomerates of stars. The study of galaxies 
is incomplete if it does not take into account the ISM, which for late--type 
galaxies accounts for more mass than that of stars. Besides, it is expected that
in the deep past, galaxies were gas--dominated and with the passing of time the cold 
gas was being transformed into stars. The ISM is a turbulent, non--isothermal,
multi--phase flow. Most of the gas mass is contained in neutral instable
HI clouds ($10^2<T< 10^4$K) and in dense, cold molecular clouds ($T<10^2$K), 
where stars form. Most of the volume of the ISM is occuppied by diffuse 
($n\approx 0.1$cm$^{-3}$), warm--hot ($T\approx 10^4-10^5$K) turbulent gas
that confines clouds by pressure.
The complex structure of the ISM is related to (i) its peculiar
thermodynamical properties (in particular the heating and cooling processes),
(ii) its hydrodynamical and magnetic properties which imply development of 
turbulence, and (iii) the different energy input sources. The star formation
unities (molecular clouds) appear to form during large--scale compression of the 
diffuse ISM driven by supernovae (SN), magnetorotational instability, or disk 
gravitational 
instability (e.g., \cite{ballesteros}). At the same time, the energy input by stars 
influences the hydrodynamical conditions of the ISM: the star formation 
results self--regulated by a delicate energy (turbulent) balance.

Galaxies are true ``ecosystems'' where stars form, evolve and collapse 
in constant interaction with the complex ISM. Following a pedagogical
analogy with biological sciences, we may say that the study of galaxies 
proceeded through taxonomical, anatomical, ecological and genetical
approaches.

\subsection{Taxonomy} %

As it happens in any science, as soon as galaxies were discovered,  the next step
was to attempt to classify these news objects. This endeavor was taken on by 
E. Hubble. 
The showiest characteristics of galaxies are the bright shapes produced
by their stars, in particular those most luminous. Hubble noticed that
by their external look (morphology), galaxies can be divided 
into three principal types: Ellipticals (E, from round to flattened elliptical shapes),
Spirals (S, characterized by spiral arms emanating from their central regions
where an spheroidal structure called bulge is present), 
and Irregulars (Irr, clumpy without any defined shape). In fact, the last two 
classes of galaxies are disk--dominated, rotating structures. Spirals are subdivided
into Sa, Sb, Sc types according to the size of the bulge in relation to the disk,
the openness of the winding of the spiral arms, and the degree of resolution
of the arms into stars (in between the arms there are also stars but less luminous
than in the arms). Roughly 40\% of S galaxies present an extended rectangular 
structure (called  bar) further from the bulge; these are the barred Spirals 
(SB), where the bar is evidence of disk gravitational instability.

From the physical point of view, the most remarkable aspect of the morphological 
Hubble sequence is the ratio of spheroid (bulge) to total luminosity. This ratio
decreases from 1 for the Es, to $\sim 0.5$ for the so--called lenticulars (S0),
to $\sim 0.5-0.1$ for the Ss, to almost 0 for the Irrs. {\it What is the origin
of this sequence? Is it given by nature or nurture? Can the morphological
types change from one to another and how frequently they do it?} It is interesting
enough that roughly half of the stars at present are in galaxy spheroids
(Es and the bulges of S0s and Ss), while the other half is in disks
(e.g., \cite{Bell04}), where some fraction of stars is still forming.

\subsection{Anatomy}

The morphological classification of galaxies is based on their external
aspect and it implies somewhat subjective criteria. Besides, the 
``showy'' features that characterize this classification may change
with the color band: in  blue bands, which trace young luminous
stellar populations, the arms, bar and other features may look 
different to what it is seen in infrared bands, which trace less
massive, older stellar populations. We would like to explore deeper the
internal physical properties of galaxies and see whether these
properties correlate along the Hubble sequence. Fortunately, this
seems to be the case in general so that, in spite of the complexity
of galaxies, some clear and sequential trends in their properties 
encourage us to think about regularity and the possibility to find
driving parameters and factors beyond this complexity. 

\begin{figure}
\centering
\includegraphics[height=10cm]{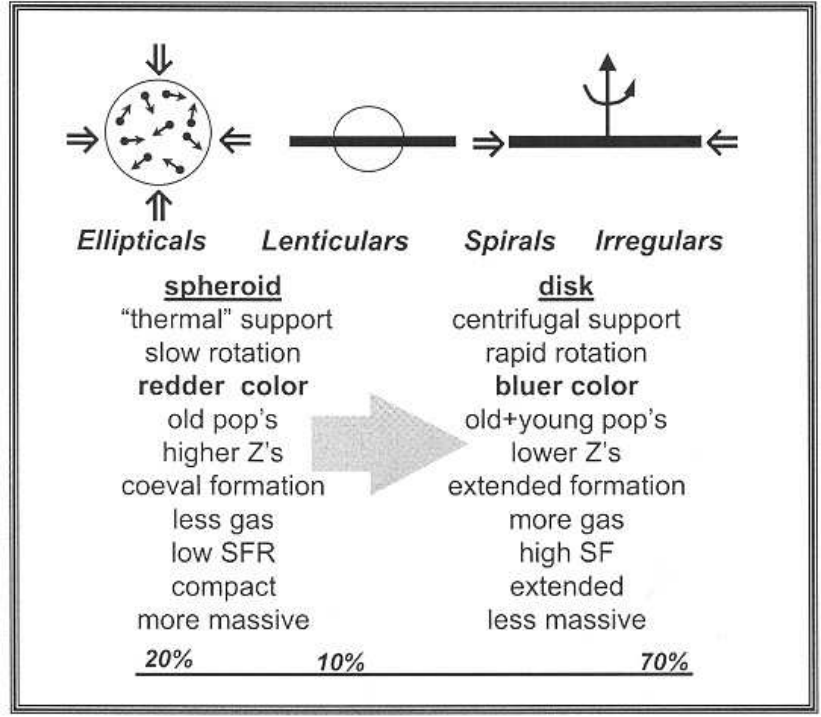}
%
%
\caption{Main trends of physical properties of galaxies along the Hubble 
morphological sequence. The latter is basically a sequence of change
of the spheroid--to--disk ratio. Spheroids are supported against gravity
by velocity dispersion, while disks by rotation.}
\label{galprop}       
\end{figure}

Figure \ref{galprop} below resumes the main trends of the ``anatomical''
properties of galaxies along the Hubble sequence.  

The advent of extremely large galaxy surveys made possible massive and uniform 
determinations of global galaxy properties. 
Among others, the Sloan Digital Sky Survey (SDSS\footnote
{\textit{www.sdss.org/sdss.html}}) and the Two--degree Field Galaxy Redshift 
Survey (2dFGRS\footnote{\textit{www.aao.gov.au/2df/}})  
currently provide uniform data already for around $10^5$ galaxies in 
limited volumes. The numbers will continue growing in the coming years.
The results from these surveys confirmed the well known trends shown in 
Fig. \ref{galprop}; moreover, it allowed to determine the distributions
of different properties. Most of these properties present a {\it bimodal}
distribution with two main sequences: the red, passive galaxies
and the blue, active galaxies, with a fraction of intermediate types
(see for recent results \cite{Kauffmann04,Balogh04,Tanaka04, 
Croton05,Weinmann06} and more references therein). The most distinct
segregation in two peaks is for the specific star formation rate
($\dot{M_s}/M_s$); there is a narrow and high peak of passive galaxies,
and a broad and low peak of star forming galaxies. The two sequences
are also segregated in the luminosity
function: the faint end is dominated by the blue, active sequence,
while the bright end is dominated by the red, passive sequence.
It seems that the transition from one sequence to the other
happens at the galaxy stellar mass of $\sim 3\times 10^{10}\msun$.

\paragraph{The hidden component}

\begin{figure}
\centering
\includegraphics[height=9cm]{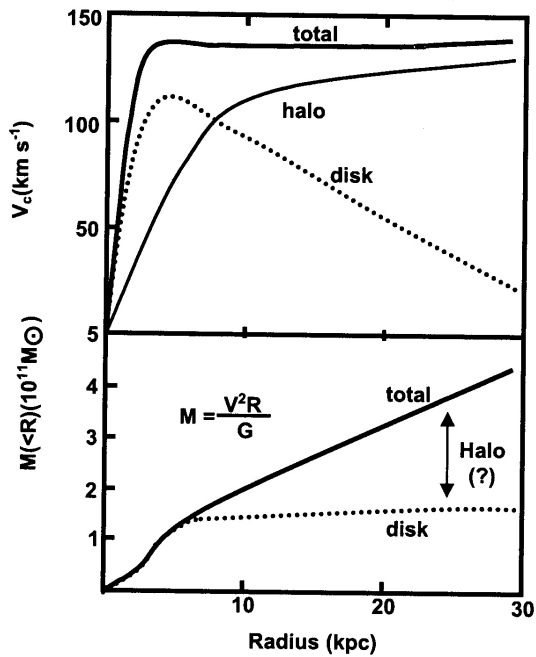}
%
%
\caption{Under the assumption of circular orbits, the observed rotation curve
of disk galaxies traces the dynamical (total) mass distribution. The outer rotation
curve of a nearly exponential disk decreases as in the Keplerian case. The
observed rotation curves are nearly flat, suggesting the existence of massive
dark halos.}
\label{rotcur}       
\end{figure}

Under the assumption of Newtonian gravity, the observed dynamics of galaxies
points out to the presence of enormous amounts of mass not seen as stars or
gas. Assuming that disks are in centrifugal equilibrium and that the 
orbits are circular (both are reasonable assumptions for non--central 
regions), the measured rotation curves are good tracers of the total
(dynamical) mass distribution (Fig. \ref{rotcur}). The mass distribution
associated with the luminous galaxy (stars+gas) can be inferred directly
from the surface brightness (density) profiles.  For an exponential
disk of scalelength $R_d$ (=3 kpc for our Galaxy), the rotation curve beyond 
the optical radius ($R_{opt}\approx 3.2R_d$) decreases as in the Keplerian case. 
The observed
HI rotation curves at radii around and beyond $R_{opt}$ are far from
the Keplerian fall--off,  implying the existence of hidden mass called
{\it dark matter (DM)} \cite{rubin,bosma}. The fraction of DM increases with 
radius.

It is important to remark the following observational facts:

{\narrower
$\bullet$ the {\it outer rotation curves are not universally flat} as it is assumed
in hundreds of papers. Following, Salucci \& Gentile \cite{salucci05},
let us define the average value of the rotation curve logarithmic slope,
$\bigtriangledown\equiv (dlog V/dlog R)$ between two and three $R_d$. A flat curve
means $\bigtriangledown= 0$; for an exponential disk without DM,  $\bigtriangledown=- 0.27$ at 3$R_s$. 
Observations show a large range of values for the slope: $-0.2\le\bigtriangledown\le 1$

$\bullet$ the rotation curve shape ($\bigtriangledown$) correlates with the luminosity and 
surface brightness of galaxies \cite{persic96,verheijen97,zavala}: it increases
according the galaxy is fainter and of lower surface brightness  

$\bullet$ at the optical radius $R_{opt}$, the DM--to--baryon ratio varies 
from $\approx 1$ to 7 for luminous high--surface brightness to faint
low--surface brightness galaxies, respectively     

$\bullet$ the roughly smooth shape of the rotation curves implies a fine coupling
between disk and DM halo mass distributions \cite{casertano}

\par
}

The HI rotation curves extend typically to $2-5R_{opt}$. The dynamics at larger
radii can be traced with satellite galaxies if the satellite statistics allows for
 that.  More recently, the technique of (statistical) {\it weak lensing}
 around galaxies began to emerge
as the most direct way to trace the masses of galaxy halos. The results
show that a typical $L_*$ galaxy (early or late) with a stellar mass of $M_s\approx
6\times 10^{10}$\msun\ is surrounded by a halo of $\approx 2\times 10^{12}\msun$
(\cite{mandelbaum} and more references therein). The extension of the halo
is typically $\approx 200-250$kpc. These numbers are very close to the
determinations for our own Galaxy. 

\begin{figure}
\centering
\includegraphics[height=8cm]{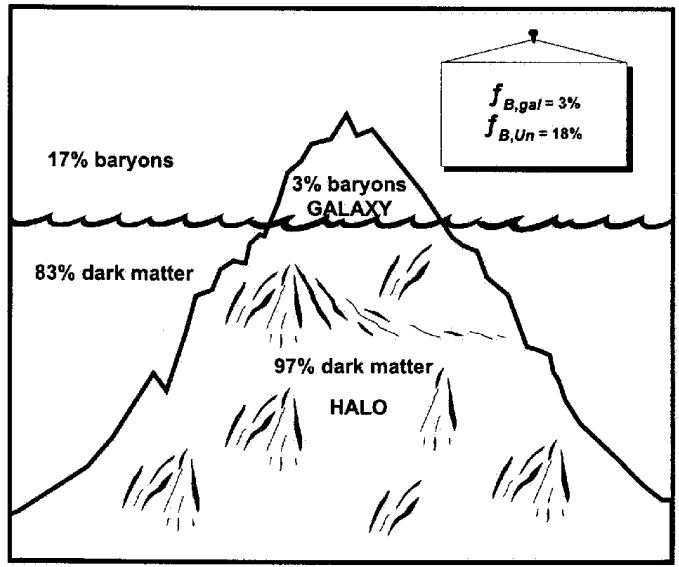}
%
%
\caption{Galaxies are just the top of the iceberg. They are surrounded by
enormous DM halos extending 10--20 times their sizes, where baryon matter is 
only less than 5\% of the total mass. Moreover, galaxies are much more 
DM--dominated than the average content of the Universe. The corresponding
typical baryon--to--DM mass ratios are given in the inset.}
\label{iceberg}       
\end{figure}

The picture has been confirmed definitively: 
luminous galaxies are just the top of the iceberg (Fig. \ref{iceberg}). The 
baryonic mass of (normal) galaxies is only $\sim 3-5\%$ 
of the DM mass in the halo! This fraction could be even lower for dwarf galaxies
(because of feedback) and for very luminous galaxies (because the gas cooling time 
$>$ Hubble time). On the other hand, the universal baryon--to--DM fraction 
($\Omega_B/\Omega_{DM}\approx 0.04/0.022$, see below) is $f_{B,Un}\approx 18\%$. 
Thus, galaxies are not only
dominated by DM, but are much more so than the average in the Universe!
This begs the next question: if the majority of baryons is not in galaxies,
where it is? Recent observations, based on highly ionized absorption lines
towards low redshfit luminous AGNs, seem to have found a fraction of the missing 
baryons in the interfilamentary warm--hot intergalactic medium at 
$T\lesssim 10^5-10^7$ K \cite{nicastro}.

\subparagraph{Global baryon inventory:} The different probes of baryon abundance
in the Universe (primordial nucleosynthesis of light elements, the ratios
of odd and even CMBR acoustic peaks heights, absorption lines in the 
Ly$\alpha$ forest) have been converging in the last years towards the same value
of the baryon density: $\Omega_b\approx 0.042\pm 0.005$. In Table 1 below, 
the densities ($\Omega'$s) of different baryon components at low redshfits 
and at $z>2$ are given (from \cite{fukugita04} and \cite{nicastro}).

\begin{table}
\centering
\caption{Abundances of the different baryon components ($h=0.7$)}
\label{tab:1}       
\begin{tabular}{lr}
\hline\noalign{\smallskip}
Component &       Contribution to $\Omega$ \\
\noalign{\smallskip}\hline\noalign{\smallskip}
{\it Low redshifts}  &         \\
Galaxies: stars      &      $0.0027\pm 0.0005$ \\
Galaxies: HI         &      ($4.2\pm 0.7$)$\times 10^{-4}$ \\
Galaxies: H$_2$      &      ($1.6\pm 0.6$)$\times 10^{-4}$ \\
Galaxies: others     &      ($\approx 2.0$)$\times 10^{-4}$ \\
Intracluster gas     &      $0.0018\pm 0.0007$ \\
IGM: (cold-warm)     &     $0.013\pm 0.0023$ \\
IGM: (warm-hot)      &     $\approx 0.016$ \\
{\it $z>2$}          &     \\
Ly$\alpha$ forest clouds   &      $>0.035$ \\
\noalign{\smallskip}\hline
\end{tabular}
\end{table}

The present--day abundance of baryons in virialized objects (normal stars, gas,
white dwarfs, black holes, etc. in galaxies,  and hot gas in clusters)  
is therefore $\Omega_B\approx 0.0037$, which accounts for $\approx 9\%$ of all 
the baryons at low redshifts. The gas in not virialized structures
in the Intergalactic Medium (cold-warm Ly$\alpha/\beta$ gas clouds and the warm--hot 
phase) accounts for $\approx 73\%$ of all baryons. Instead, at $z>2$ more
than $88\%$ of the universal baryonic fraction is in the Ly$\alpha$ forest
composed of cold HI clouds. The baryonic budget's outstanding questions: {\it Why 
only $\approx 9\%$ of baryons are in virialized structures at the present epoch?}  


\subsection{Ecology}

The properties of galaxies vary systematically as a function of environment.
The environment can be relatively local (measured through the number of
nearest neighborhoods) or of large scale (measured through counting
in defined volumes around the galaxy). 
The morphological type of galaxies is earlier in the locally 
denser regions (morphology--density relation),the fraction of ellipticals 
being maximal in cluster cores \cite{Dressler80} and enhanced in rich \cite{PG84} 
and poor groups.  The extension of the 
morphology--density relation to low local--density environment 
(cluster outskirts, low mass groups, field) has been a matter of debate. From 
an analysis of SDSS data, \cite{Goto03} have found that (i) in the sparsest regions 
both relations flatten out, (ii) in the intermediate density regions (e.g., 
cluster outskirts) the intermediate--type galaxy (mostly S0s) fraction 
increases towards 
denser regions whereas the late--type galaxy fraction decreases, and (iii) in 
the densest regions intermediate--type fraction decreases radically and early--type 
fraction increases. In a similar way, a study based on 2dFGRS data of the luminosity 
functions in clusters and voids shows that the population of faint 
late--type galaxies dominates in the latter, while, in contrast, very bright 
early--late galaxies are relatively overabundant in the former \cite{Croton05}. 
This and other studies suggest that the origin of the morphology--density 
(or morphology-radius) relation could 
be a combination of (i) {\it initial (cosmological) conditions} and (ii) of 
{\it external mechanisms} (ram-pressure and tidal stripping, thermal evaporation 
of the disk gas, strangulation, galaxy harassment, truncated star formation, etc.) 
that operate mostly in dense environments, where precisely the relation steepens 
significantly.

The morphology--environment relation evolves. It systematically flattens with $z$
in the sense that the grow of the early-type (E+S0) galaxy fraction with density 
becomes less rapid (\cite{Postman05} and more references therein) 
the main change being in the high--density population fraction. Postman et al. 
conclude that the observed flattening of the relation up to $z\sim 1$ is due 
mainly to a deficit of S0 galaxies and an excess of Sp+Irr galaxies relative to 
the local galaxy population; the E fraction-density relation does not appear
to evolve over the range $0<z<1.3$!
Observational studies show that other properties besides morphology vary 
with environment. The galaxy properties 
most sensitive to environment are the integral color and specific star formation
rate (e.g. \cite{Kauffmann04,Tanaka04,Weinmann06}. The dependences of both properties 
on environment extend typically to lower densities than the dependence for morphology. 
These properties are tightly related to the galaxy star formation history, 
which in turn depends on internal formation/evolution processes related 
directly to initial cosmological conditions 
as well as to external astrophysical mechanisms able to inhibit or 
induce star formation activity.

\subsection{Genetics}

Galaxies definitively evolve. We can reconstruct the past of a given
galaxy by matching the observational properties of its stellar
populations and ISM with (parametric) spectro--photo--chemical models
(inductive approach). 
These are well--established models specialized in following the spectral, 
photometrical and chemical evolution of stellar populations formed 
with different gas infall rates and star formation laws (e.g.
\cite{boissier} and the references therein). The inductive
approach allowed  to determine that spiral galaxies as our
Galaxy can not be explained with closed--box models (a single burst
of star formation); continuous infall of low--metallicity gas
is required to reproduce the local and global colors, metal abundances,
star formation rates, and gas fractions. On the other hand, the properties
of massive ellipticals (specially their high $\alpha$-elements/Fe ratios) 
are well explained by a single early fast burst of star formation and subsequent 
passive evolution. 

A different approach to the genetical study of galaxies emerged after cosmology
provided a reliable theoretical background.
Within such a background it is possible to ``handle'' galaxies as physical 
objects that evolve according to the initial and boundary conditions given
by cosmology. The deductive construction of galaxies can be confronted
with observations corresponding to different stages of the proto-galaxy
and galaxy evolution. The breakthrough for the deductive approach
was the success of the inflationary theory and the consistency of 
the so--called Cold Dark Matter (CDM) scenario with particle physics and observational 
cosmology. The main goal of these notes is to describe the ingredients, predictions,
and tests of this scenario.

\paragraph{Galaxy evolution in action}

The dramatic development of observational astronomy in the last 15 years or so
opened a new window for the study of galaxy genesis: the follow up of 
galaxy/protogalaxy populations and their environment at different redshifts.
The Deep and Ultra Deep Fields of the Hubble Spatial Telescope and other
facilities allowed to discover new populations of galaxies 
at high redshifts, as well as to measure the evolution of global (per unit 
of comoving volume) quantities associated with galaxies: the cosmic
star formation rate density (SFRD), the cosmic density of neutral gas, the
cosmic density of metals, etc. Overall, these global quantities
change significantly with $z$, in particular the SFRD as traced by
the UV--luminosity at rest of galaxies \cite{madau}: since $z\sim 1.5-2$ to the present
it decreased by a factor close to ten (the Universe is literally lightening off), 
and for higher redshifts the SFRD remains roughly constant or slightly decreases
(\cite{giavalisco04,hopkins} and the references therein). There exists indications 
that the SFRD at redshifts 2--4 could be approximately two times
higher if considering Far Infrared/submmilimetric sources (SCUBA
galaxies), where intense bursts of star formation take place in a
dust--obscured phase.       

Concerning populations of individual galaxies, the Deep Fields evidence
a significant increase in the fraction of blue galaxies at $z\sim 1$ for the 
blue sequence that at these epochs look more distorted and with higher
SFRs than their local counterparts. Instead, the changes observed in the red 
sequence are 
small; it seems that most red elliptical galaxies were in place long ago.
At higher redshifts ($z\grtsim 2$), galaxy objects with high SFRs become more 
and more common. The most abundant populations are:

\subparagraph{Lyman Break Galaxies (LBG)}, selected via the Lyman break
at 912\AA~ in the rest--frame. These are star--bursting galaxies (SFRs
of $10-1000$\msun/yr) with stellar masses of $10^{9}- 10^{11}$\msun\
and moderately clustered.

\subparagraph{Sub-millimeter (SCUBA) Galaxies,} detected
with sub--millimeter bolometer arrays. These are strongly star--bursting
galaxies (SFRs of $\sim 1000$\msun/yr) obscured by dust; they are
strongly clustered and seem to be merging galaxies, probably precursors
of ellipticals. 

\subparagraph{Lyman $\alpha$ emitters (LAEs),} selected in narrow--band
studies centered in the Lyman $\alpha$ line at rest at $z>3$; strong
emission Lyman $\alpha$ lines evidence phases of rapid star formation 
or strong gas cooling. LAEs could be young (disk?) galaxies in the 
early phases of rapid star formation or even before, when the gas 
in the halo was cooling and infalling to form the gaseous disk. 

\subparagraph{Quasars (QSOs),} easily discovered by their powerful
energetics; they are associated to intense activity in the nuclei
of galaxies that apparently will end as spheroids; QSOs are strongly
clustered and are observed up to $z\approx 6.5$.

There are many other populations of galaxies and protogalaxies at high redshifts
(Luminous Red Galaxies, Damped Ly$\alpha$ disks, Radiogalaxies, etc.).
A major challenge now is to put together all the pieces of the high--redshift
puzzle to come up with a coherent picture of galaxy formation and
evolution.

\section{Cosmic structure formation}

In the previous section we have learn that galaxy formation and
evolution are definitively related to cosmological conditions.
Cosmology provides the theoretical framework for the 
initial and boundary conditions of the cosmic structure formation
models. At the same time, the confrontation of model predictions
with astronomical observations became the most powerful testbed 
for cosmology. As a result of this fruitful convergence between 
cosmology and astronomy, there emerged the current paradigmatic scenario 
of cosmic structure formation and evolution of the Universe 
called $\Lambda$ Cold Dark Matter (\Lcdm). The \Lcdm\ scenario
integrates nicely: (1) cosmological theories (Big Bang and Inflation),
(2) physical models (standard and extensions of the particle physics models),
(3) astrophysical models (gravitational cosmic structure growth, hierarchical 
clustering, gastrophysics), and (4) phenomenology (CMBR anisotropies,
non-baryonic DM, repulsive dark energy, flat geometry, galaxy properties).

Nowadays, cosmology passed from being the Cinderella of astronomy 
to be one of the highest precision sciences. Let us consider only the 
Inflation/Big Bang cosmological models with the F-R-W
metric and adiabatic perturbations. The number of parameters that
characterize these models is high, around 15 to be more precise.
No single cosmological probe constrain all of these parameters. By using 
 multiple data sets and probes it is possible to constrain with precision
several of these parameters, many of which correlate among them (degeneracy). 
The main cosmological probes used for
precision cosmology are the CMBR anisotropies, the type--Ia SNe and long
Gamma--Ray Bursts, 
the Ly$\alpha$ power spectrum, the large--scale power spectrum from
galaxy surveys, the cluster of galaxies dynamics and abundances, the  
peculiar velocity surveys, the weak and strong lensing,
the baryonic acoustic oscillation in the large--scale galaxy distribution.
There is a model that is systematically consistent with most of these probes
and one of the goals in the last years has been to improve the error
bars of the parameters for this 'concordance' model. The geometry in the
concordance model is flat with an energy composition dominated in $\sim 2/3$ by 
the cosmological constant $\Lambda$ (generically called Dark Energy),
responsible for the current accelerated expansion of the Universe. The other
$\sim 1/3$ is matter, but $\sim 85\%$ of this 1/3 is in form of 
non--baryonic DM. Table 2 presents the central values of  
different parameters of the \Lcdm\ cosmology from combined model fittings 
to the recent 3--year $WMAP$ CMBR and several other cosmological
probes \cite{spergel06} (see the WMAP website).

\begin{table}
\centering
\caption{Constraints to the parameters of the \Lcdm\ model}
\label{tab:1}       
%
%
\begin{tabular}{lr}
\hline\noalign{\smallskip}
Parameter &   Constraint\\
\noalign{\smallskip}\hline\noalign{\smallskip}
Total density       & $\Omega = 1$ \\
Dark Energy density & $\Omega_{\Lambda}=0.74$ \\
Dark Matter density & $\Omega_{DM}=0.216$ \\
Baryon Matter dens. & $\Omega_{B}=0.044$ \\
Hubble constant     & $h=0.71$ \\
Age                 & 13.8 Gyr \\
Power spectrum norm. & $\sigma_8=0.75$ \\
Power spectrum index & $n_s(0.002)=0.94$ \\
\noalign{\smallskip}\hline
\end{tabular}
\end{table}

In the following, I will describe some of the ingredients of the \Lcdm\ scenario,
emphasizing that most of these ingredients are well established aspects
that any alternative scenario to \Lcdm\ should be able to explain.

\subsection{Origin of fluctuations}

The Big Bang\footnote{It is well known that the name of 'Big Bang' 
is not appropriate for this theory. The key physical conditions required
for an explosion are temperature and pressure gradients. These
conditions contradict the Cosmological Principle of
homogeneity and isotropy on which is based the 'Big Bang' theory.} 
is now a mature theory, based on well established
observational pieces of evidence. 
However, the Big Bang theory has limitations. One of them
is namely the origin of fluctuations that should give rise
to the highly inhomogeneous structure observed today in the
Universe, at scales of less than $\sim 200$Mpc. The smaller
the scales, the more clustered is the matter. For example,
the densities inside the central regions of galaxies, within
the galaxies, cluster of galaxies, and superclusters are about
$10^{11}$, $10^6$, $10^3$ and few times the average density
of the Universe, respectively. 

The General Relativity equations that describe the Universe dynamics
in the Big Bang theory are for an homogeneous and isotropic fluid 
(Cosmological Principle); inhomogeneities are not taken into 
account in this theory ``by definition''.  Instead, the concept
of fluctuations is inherent to the Inflationary theory introduced
in the early 80's by A. Guth and A. Linde namely to overcome the 
Big Bang limitations. According to this theory, at the energies
of Grand Unification ($\grtsim 10^{14}$GeV or $T\grtsim 10^{27}$K!),
the matter was in the state known in quantum field theory as vacuum.
Vacuum is characterized by quantum fluctuations --temporary changes 
in the amount of energy in a point in space, arising from Heisenberg 
uncertainty principle. For a small time interval $\Delta t$, a virtual
particle--antiparticle pair of energy $\Delta E$ is created (in the 
GU theory, the field particles are supposed to be the X- and Y-bossons),
but then the pair disappears so that there is no violation of energy 
conservation. Time and energy are related by 
$\Delta E \Delta t \approx {h \over 2 \pi}$. The vacuum quantum 
fluctuations are proposed to be the seeds of present--day 
structures in the Universe. 

How is that quantum fluctuations become density inhomogeneities? 
During the inflationary period, the expansion is described approximately
by the de Sitter cosmology, $a\propto e^{Ht}$, $H\equiv\dot{a}/a$ is the 
Hubble parameter and it is constant in this cosmology. Therefore, the proper length of 
any fluctuation grows as $\lambda_p\propto e^{Ht}$. On the other hand, the 
proper radius of the horizon for de Sitter metric is equal to $c/H=$const,
so that initially causally connected (quantum) fluctuations become 
suddenly supra--horizon (classical)  perturbations to the spacetime 
metric. After inflation, 
the Hubble radius grows proportional to $ct$, and at some time a given curvature
perturbation cross again the horizon (becomes causally connected,
$\lambda_p<L_H$). It becomes now a true density perturbation. The interesting
aspect of the perturbation 'trip' outside the horizon is that its amplitude
remains roughly constant, so that if the amplitude of the fluctuations
at the time of exiting the horizon during inflation is constant (scale invariant), 
then their amplitude at the time of entering the horizon should be also scale 
invariant. In fact, the computation of classical perturbations generated 
by a quantum field during inflation demonstrates that the amplitude of the scalar 
fluctuations at the time of crossing the horizon is nearly constant, 
$\delta\phi_{H}\propto$const. This can be understood on dimensional grounds:
due to the Heisenberg principle $\delta\phi/\delta t\propto$ const,
where $\delta t\propto H^{-1}$. Therefore, $\delta\phi_H\propto H$, but $H$
is roughly constant during inflation, so that $\delta\phi_H\propto$const.

\subsection{Gravitational evolution of fluctuations}

The \Lcdm\ scenario assumes the gravitational instability paradigm:
the cosmic structures in the Universe were formed as a consequence of 
the growth of primordial tiny fluctuations (for example seeded in the 
inflationary epochs) by gravitational instability in an expanding frame.
The fluctuation or perturbation is characterized by its density contrast,
\begin{equation}
\delta\equiv \frac{\delta\rho}{\overline{\rho}}= \frac{\rho - \overline{\rho}}
{\overline{\rho}}, 
\end{equation}
where $\overline{\rho}$ is the average density of the Universe and 
$\rho$ is the perturbation density. At early epochs, $\delta << 1$
for perturbation of all scales, otherwise the homogeneity condition
in the Big Bang theory is not anymore obeyed. When $\delta << 1$, the
perturbation is in the {\it linear} regime and its physical size grows
with the expansion proportional to $a(t)$. The perturbation
analysis in the linear approximation shows whether a given perturbation 
is stable ($\delta\sim$ const or even $\rightarrow 0$) or unstable ($\delta$ grows). 
In the latter case, when $\delta\rightarrow 1$, the linear approximation
is not anymore valid, and the perturbation ``separates'' from the expansion,
collapses, and becomes a self--gravitating structure. The gravitational
evolution in the {\it non--linear regime} is complex for realistic
cases and is studied with numerical N--body simulations. Next,
a pedagogical review of the linear evolution of perturbations is presented.
More detailed explanations on this subject can be found in the books 
\cite{KT,peebles93,padma,coles,Longairbook,Peacock}.

\paragraph{Relevant times and scales.} 

The important times in the problem of linear gravitational evolution of 
perturbations are: (a) the epoch when inflation 
finished ($t_{inf}\approx 10^{-34}$s, at this time the primordial fluctuation 
field is established); (b) the epoch of matter--radiation equality \te\ 
(corresponding to $\ae\approx 1/3.9\times10^4(\Omega_0 h^2)$, before
\te\ the dynamics of the universe is dominated by radiation density,
after \te\ dominates matter density); (c) the epoch of recombination \trec, 
when radiation decouples from baryonic matter (corresponding
to $\arec=1/1080$, or $\trec\approx 3.8\times 10^{5}$yr for the concordance
cosmology). 

Scales: first of all, we need to characterize
the size of the perturbation. In the linear regime, its physical size
expands with the Universe: $\lp = a(t) \lambda_0$, where $\lambda_0$ is the 
comoving size,  by convention fixed (extrapolated) to the present epoch, 
$a(t_0)=1$. In a given (early) epoch, the size of the perturbation can be 
larger than the so--called {\it Hubble radius}, the typical radius over which 
physical processes operate coherently (there is causal connection): 
$\lh\equiv(a/\dot{a})^{-1}= H^{-1}=n^{-1} ct$. For the radiation or matter
dominated cases, $a(t)\propto t^n$, with $n=1/2$ and $n=2/3$, respectively,
that is $n<1$. Therefore, $L_H$ grows faster than \lp\ and at a given ``crossing'' 
time \tcr, $\lp<\lh$. Thus, the 
perturbation is supra--horizon sized at epochs $t<\tcr$ and sub--horizon 
sized at $t>\tcr$. Notice that if $n>1$, then at some time the 
perturbation ``exits'' the Hubble radius. This is what happens in the 
inflationary epoch, when $a(t)\propto e^{t}$: causally--connected
fluctuations of any size are are suddenly ``taken out'' outside the 
Hubble radius becoming causally disconnected.

For convenience, in some cases it is better to use masses instead of sizes.
Since in the linear regime $\delta<<1$ ($\rho\approx \overline{\rho}$),
then $M\approx \rho_M(a)\ell^3$, where $\ell$ is the size of a 
given region of the Universe with average matter density $\rho_M$. The
mass of the perturbation, $M_p$, is invariant. 

\paragraph{Supra--horizon sized perturbations.}

In this case, causal, microphysical processes are not possible, so that it
does not matter what perturbations are made of (baryons, radiation, dark matter,
etc.); they are in general just perturbations to the metric. To study the 
gravitational growth of metric perturbations, a General Relativistic 
analysis is necessary. A major issue in carrying out this program is that 
the metric perturbation is not a gauge invariant quantity. See e.g.,
\cite{KT} for an outline of how E. Lifshitz resolved brilliantly this difficult
problem in 1946. The result is quite simple and it shows that 
the amplitude of metric perturbations outside the horizon grows
{\it kinematically} at different rates, depending on the dominant 
component in the expansion dynamics. For the critical cosmological
model (at early epochs all models approach this case), the growing
modes of metric perturbations according to what dominates the background are:
\begin{eqnarray}
\delta_{m,+}\propto a(t)\propto t^{2/3},.................matter \\ \nonumber
\delta_{m,+}\propto a(t)^2\propto t,.................radiation \\
\delta_{m,+}\propto a(t)^{-2}\propto e^{-2Ht},..\Lambda\ (de Sitter) 
\end{eqnarray}   

\paragraph{Sub--horizon sized perturbations.} 

Once perturbations are causally connected, microphysical processes are
switched on (pressure, viscosity, radiative transport, etc.)
and the gravitational evolution of the perturbation depends
on what it is made of. Now, we deal with true {\it density} perturbations.
For them applies the classical perturbation analysis for a fluid, originally 
introduced by J. Jeans in 1902, in the context of the problem of star formation 
in the ISM.  
But unlike in the ISM, in the cosmological context the fluid
is expanding.  What can prevent the perturbation amplitude from growing
gravitationally? The answer is pressure support. If the fluid pressure 
gradient can re--adjust itself in a timescale \tp\ smaller than the 
gravitational collapse timescale, \tg, then pressure prevents the gravitational 
growth of $\delta$. Thus, the condition for gravitational instability is:
\begin{equation}
\tg\approx \frac{1}{(G\rho)^{1/2}} < \tp\approx \frac{\lp}{v},
\end{equation}
where $\rho$ is the density of the component that is most
gravitationally dominant in the Universe, and $v$ is the sound speed 
(collisional fluid) or velocity dispersion (collisionless fluid) of the 
perturbed component. In other words, if the perturbation scale is larger than
a critical scale $\lambda_J\sim v(G\rho)^{-1/2}$, then pressure
loses, gravity wins. 

The perturbation analysis applied to the
hydrodynamical equations of a fluid at rest shows that $\delta$
grows {\it exponentially} with time for perturbations obeying the 
Jeans instability criterion $\lp>\lambda_J$, where the
exact value of $\lambda_J$ is $v(\pi/G\rho)^{1/2}$. If $\lp<\lambda_J$,
then the perturbations are described by stable {\it gravito--acustic oscillations}.
The situation is conceptually similar for perturbations in an expanding 
cosmological fluid, but the growth of $\delta$ in the unstable
regime is {\it algebraical} instead of exponential. Thus, the cosmic 
structure formation process is relatively slow. Indeed, the typical
epochs of galaxy and cluster of galaxies formation are at redshifts 
$z\sim 1-5$ (ages of $\sim 1.2-6$ Gyrs) and $z<1$ (ages larger
than 6 Gyrs), respectively.

\subparagraph{Baryonic matter.} The Jeans instability analysis for a 
relativistic (plasma) fluid of baryons {\it ideally} 
coupled to radiation and expanding at the rate $H=\dot{a}/a$ shows that
there is an instability critical scale 
$\lambda_J = v(3\pi/8G\rho)^{1/2}$, where the sound speed for 
adiabatic perturbations is $v=p/\rho=c/\sqrt{3}$; the latter
equality is due to pressure radiation. At the epoch when 
{\it radiation dominates}, $\rho=\rho_r\propto a^{-4}$ and then 
$\lambda_J\propto a^2\propto ct$. It is not surprising that at this 
epoch $\lambda_J$ approximates the Hubble scale $\lh\propto ct$ 
(it is in fact $\sim 3$ times larger). Thus, perturbations that might
collapse gravitationally are in fact outside the horizon, and those
that already entered the horizon, have scales smaller than $\lambda_J$:
they are stable gravito--acoustic oscillations. When {\it matter
dominates}, $\rho=\rho_M\propto a^{-3}$, and $a\propto t^{2/3}$. 
Therefore,  $\lambda_J\propto a\propto t^{2/3} \lesssim \lh$, but still
radiation is coupled to baryons, so that radiation pressure is dominant
and $\lambda_J$ remains large. However, when radiation decouples
from baryons at \trec, the pressure support drops dramatically by
a factor of $P_r/P_b\propto n_rT/n_bT\approx 10^8$! Now, the Jeans
analysis for a gas mix of H and He at temperature 
$T_{\rm rec}\approx 4000$ K shows that baryonic clouds with masses 
$\grtsim 10^6\msun$ can collapse gravitationally, i.e. all masses of
cosmological interest. But this is literally too ``ideal'' to be true.

The problem is that as the Universe expands, radiation cools ($T_r=T_0a^{-1}$) 
and the photon--baryon fluid becomes less and less perfect: the mean free
path for scattering of photons by electrons (which at the same time
are coupled electrostatically to the protons) increases. Therefore, photons can
diffuse out of the bigger and bigger density perturbations as the photon
mean free path increases. If perturbations are in the gravito--acoustic 
oscillatory regime, then the oscillations are damped out and the
perturbations disappear. The ``ironing out'' of perturbations continues until 
the epoch of recombination. In a pioneering work, J. Silk \cite{silk68} 
carried out a perturbation analysis of a relativistic cosmological fluid 
taking into account radiative transfer in the diffusion approximation. He 
showed that all photon--baryon perturbations of masses smaller than 
$M_{S}$ are ``ironed out'' until \trec\ by the (Silk) damping process. 
The first crisis in galaxy formation theory emerged: calculations showed that
$M_{S}$ is of the order of $10^{13}-10^{14}\msunh$! If somebody 
(god, inflation, ...) seeded primordial fluctuations in the Universe,
by Silk damping all galaxy--sized perturbation are ``ironed out''.
\footnote{In the 
1970s Y. Zel'dovich and collaborators worked out a scenario of galaxy
formation starting from very large perturbations, those that were not
affected by Silk damping. In this elegant scenario, 
the large--scale perturbations, considered in a first approximation
as ellipsoids, collapse most rapidly along their shortest axis, forming
flattened structures (``pancakes''), which then fragment into galaxies
by gravitational or thermal instabilities. In this 'top-down' scenario,
to obtain galaxies in place at $z\sim 1$, the amplitude of the 
large perturbations at recombination should be $\ge 3\times 10^{-3}$. 
Observations of the CMBR anisotropies showed that the amplitudes are
1--2 order of magnitudes smaller than those required.}  

\begin{figure}
\centering
\includegraphics[height=11cm]{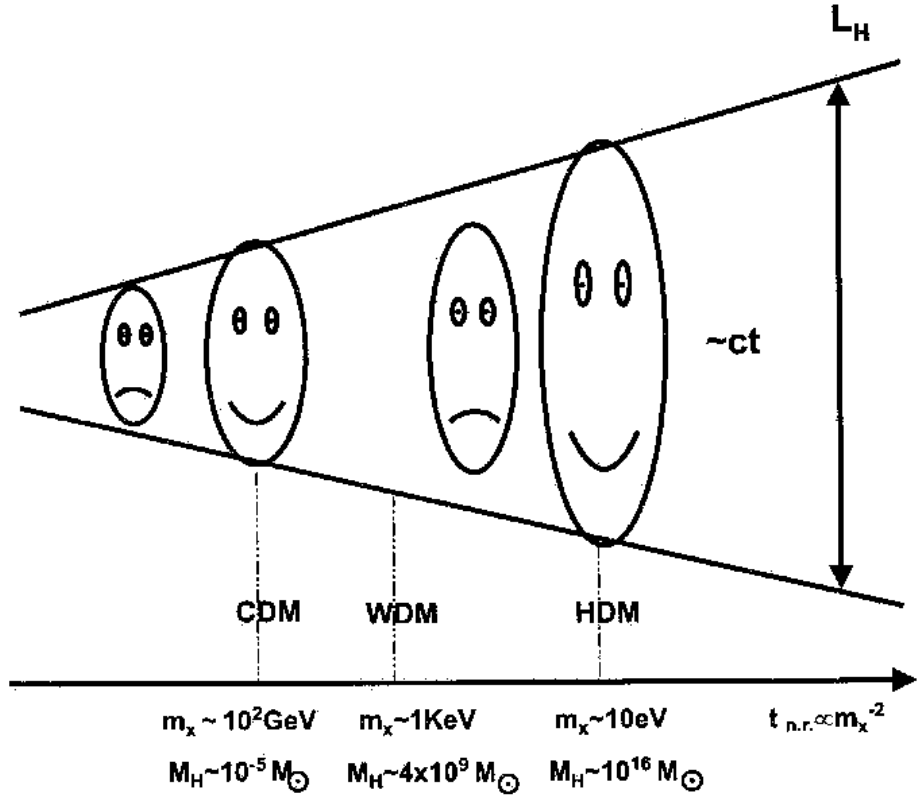}
%
%
\caption{Free--streaming damping kills perturbations of sizes roughly 
smaller than the horizon length if they are made of relativistic
particles. The epoch $t_{n.r.}$ when thermal--coupled particles become 
non--relativistic is inverse proportional to the square of the particle mass 
$m_X$. Typical particle masses of CDM, WDM and HDM are given
together with the corresponding horizon (filtering) masses.}
\label{DMfaces}       
\end{figure}

\subparagraph{Non--baryonic matter.}
The gravito--acoustic oscillations and their damping by photon diffusion
refer to baryons. What happens for a fluid of non--baryonic DM?
After all, astronomers, since Zwicky in the 1930s, find routinely pieces 
of evidence for the presence of large amounts of DM in the Universe.  
As DM is assumed to be collisionless and not interacting 
electromagnetically, then the radiative or thermal pressure supports are 
not important for linear DM perturbations. However, DM perturbations can be
damped out by {\it free streaming} if the particles are relativistic: the 
geodesic motion of the particles at the speed of light will iron out any 
perturbation smaller than a scale close to the particle horizon radius, 
because the particles can freely propagate from an overdense region to an 
underdense region. Once the particles cool and become non relativistic,
free streaming is not anymore important. A particle of mass $m_X$ and
temperature $T_X$ becomes non relativistic when $k_BT_X\sim m_Xc^2$. Since
$T_X\propto a^{-1}$, and $a\propto t^{1/2}$ when radiation dominates,
one then finds that the epoch when a thermal--relic particle becomes non 
relativistic is $t_{nr}\propto m_X^{-2}$. The more massive the DM particle, the earlier
it becomes non relativistic, and the smaller are therefore the perturbations
damped out by free streaming (those smaller than $\sim ct$; see Fig.
\ref{DMfaces}). According to the epoch when a given thermal DM particle species
becomes non relativistic, DM is called Cold Dark Matter (CDM, very early),
Warm Dark Matter (WDM, early) and Hot Dark Matter (HDM, late)\footnote{ The
reference to ``early'' and ``late'' is given by the epoch and the corresponding 
radiation temperature when the largest galaxy--sized perturbations 
($M\sim 10^{13}\msun$) enter the horizon: $a_{gal}\sim \ae\approx 1/3.9
\times10^4(\Omega_0 h^2)$ and $T_r\sim 1$KeV.}.

The only non--baryonic particles confirmed experimentally are (light) neutrinos
(HDM). For neutrinos of masses $\sim 1-10$eV,  free streaming attains to iron 
out perturbations of scales as large as massive clusters and superclusters 
of galaxies (see Fig. \ref{DMfaces}). Thus, HDM suffers the same problem
of baryonic matter concerning galaxy formation\footnote{Neutrinos exist and
have masses larger than 0.05 eV according to determinations based on solar 
neutrino oscillations. Therefore, neutrinos contribute to the matter density 
in the Universe. Cosmological observations
set a limit: $\Omega_\nu h^2 <0.0076$, otherwise too much structure is erased.}. 
At the other extreme is CDM, in which case survive free streaming practically 
all scales of cosmological interest. This makes CDM appealing to galaxy
formation theory. In the minimal CDM model, it is assumed that perturbations
of all scales survive, and that CDM particles are collisionless (they do not 
self--interact). Thus, if CDM dominates, then the first step in galaxy 
formation study is reduced to the calculation of the linear and non--linear 
gravitational evolution of collisionless CDM perturbations. Galaxies are
expected to form in the 
centers of collapsed CDM structures, called {\it halos}, from the baryonic
gas, first trapped in the gravitational potential of these halos, and
second, cooled by radiative (and turbulence) processes (see \S 5).

The CDM perturbations are free of any physical damping processes and 
in principle their amplitudes may grow by gravitational instability.
However, when radiation dominates, the perturbation growth is 
stagnated by expansion. The gravitational instability timescale for
sub--horizon linear CDM perturbations is $t_{grav}\sim (G\rho_{DM})^{-2}$,
while the expansion (Hubble) timescale is given by 
$t_{exp}\sim (G\overline{\rho})^{-2}$. When radiation dominates,
$\overline{\rho}\approx \rho_r$ and $\rho_r>>\rho_M$. Therefore
$t_{exp}<<t_{grav}$, that is, expansion is faster than the 
gravitational shrinking. 

\begin{figure}
\centering
\includegraphics[height=9.3cm]{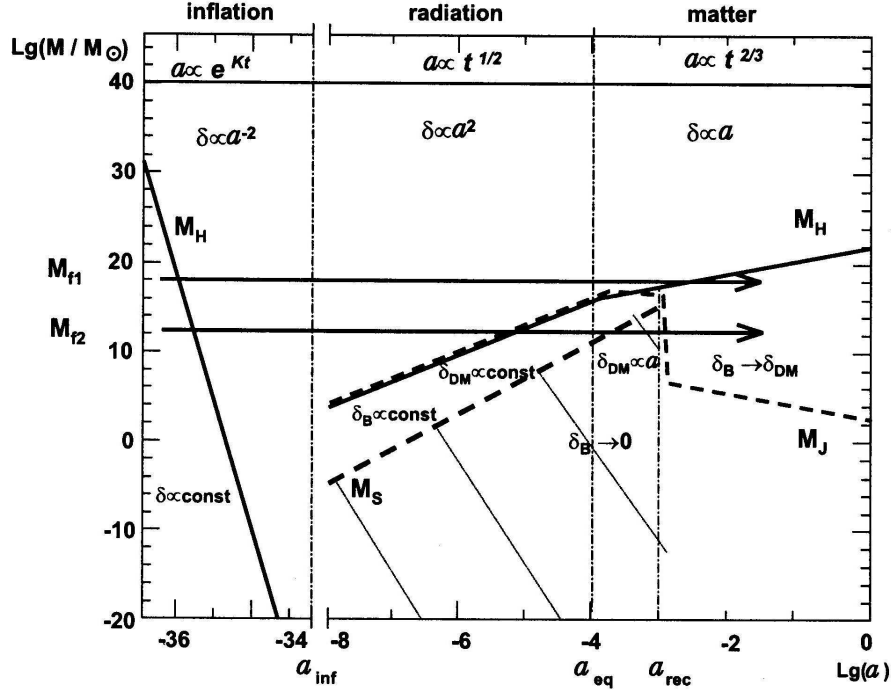}
\caption{Different evolutive regimes of perturbations $\delta$. The suffixes 
``B'' and ``DM'' are for baryon--photon and DM perturbations, respectively. 
The evolution of the horizon, Jeans and Silk masses ($M_H, M_J$, and $M_S$) are showed. 
$M_{f1}$ and $M_{f2}$ are the masses of two perturbations. See text for 
explanations.}
\label{resumen}       
\end{figure}

 Fig. \ref{resumen} resumes the evolution of primordial perturbations.
Instead of spatial scales, in Fig. \ref{resumen} are shown masses, which
are invariant for the perturbations. We highlight the following conclusions
from this plot: (1) Photon--baryon perturbations of masses $< M_S$ are washed
out ($\delta_B\rightarrow 0$) as long as baryon matter is coupled to radiation. 
(2) The amplitude of CDM perturbations that enter the horizon before 
$\te$ is ``freezed-out'' ($\delta_{DM}\propto$const) 
as long as radiation dominates; these are perturbations of masses smaller than 
$M_{H,eq}\approx 10^{13}(\Omega_M h^2)^{-2}\msun$, namely galaxy scales. (3) The
baryons are trapped gravitationally by CDM perturbations, and within a 
factor of two in $z$, baryon perturbations attain amplitudes half that of
$\delta_{DM}$. For WDM or HDM  perturbations, the free--streaming damping 
introduces a mass scale $M_{fs}\approx M_{H,n.r.}$ in Fig. \ref{resumen}, below
which $\delta\rightarrow 0$; $M_{fs}$ increases as the DM mass particle decreases
(Fig. \ref{DMfaces}). 

\subparagraph{The processed power spectrum of perturbations.}       

The exact solution to the problem of linear evolution of cosmological 
perturbations is much more complex than the conceptual aspects described 
above. Starting from a primordial fluctuation field, the perturbation 
analysis should be applied to a cosmological mix of
baryons, radiation, neutrinos, and other non--baryonic dark matter 
components (e.g., CDM), at sub-- and supra--horizon scales (the fluid
assumption is relaxed). Then, coupled 
relativistic hydrodynamic and Boltzmann equations in a general relativity
context have to be solved taking into account radiative and dissipative 
processes. The outcome of these complex calculations is the full description
of the processed fluctuation field at the recombination epoch (when
fluctuations at almost all scales are still in the linear regime). The
goal is double and of crucial relevance in cosmology and astrophysics:
{\it 1) to predict the physical and statistical properties of CMBR anisotropies,
which can be then compared with observations, and 2) to provide the initial 
conditions for calculating the non--linear regime of cosmic structure formation 
and evolution}. Fortunately,
there are now several public friendly-to-use codes that numerically
solve the cosmological linear perturbation equations (e.g., CMBFast and CAMB
\footnote{\textit{http://www.cmbfast.org} and  \textit{http://camb.info/}}).

The description of the density fluctuation field is statistical. As any
random field, it is convenient to study perturbations in the Fourier space.
The Fourier expansion of $\deltaX$ is:
\begin{eqnarray}
\deltaX=\frac{V}{(2\pi)^3}\int{\dk e^{-i\bf{kx}}d^3k}, \\
\dk=V^{-1}\int{\deltaX e^{i\bf{kx}}d^3x}
\end{eqnarray}  
The Fourier modes \dk\ evolve independently while the perturbations are in
the linear regime, so that the perturbation analysis can be applied to this
quantity. For a Gaussian random field, any statistical 
quantity of interest can be specified in terms of the power spectrum
$P(k)\equiv |\dk|^2$, which measures the amplitude of the fluctuations at 
a given scale $k$\footnote{The phases of the Fourier modes in the
Gaussian case are {\it uncorrelated}. Gaussianity is the simplest assumption
for the primordial fluctuation field statistics and it seems to be consistent
with some variants of inflation. However, there are other variants that predict
non--Gaussian fluctuations (for a recent review on this subject see e.g. 
\cite{Bartolo04}), and the observational determination of the primordial 
fluctuation statistics is currently an active field of investigation. The 
properties of cosmic structures depend on the assumption about the primordial
statistics, not only at large scales but also at galaxy scales; see for
a review and new results \cite{AR03}.}. Thus, from the linear perturbation analysis 
we may follow the evolution of $P(k)$. A more intuitive quantity than $P(k)$ is 
the mass variance $\sms\equiv\langle(\delta M/M)^2_R\rangle$ of the fluctuation field.
The physical meaning of \sm\ is that of an $rms$ density contrast on a given
sphere of radius $R$ associated to the mass $M=\rho V_W(R)$, where $W(R)$
is a window (smoothing) function. The mass variance is related to $P(k)$. By
assuming a power law power spectrum, $P(k)\propto k^n$, it is easy to 
show that 
\begin{eqnarray}
\sm\propto R^{-(3+n)}\propto M^{-(3+n)/3}=M^{-2\alpha} \\ \nonumber
\alpha=\frac{3+n}{6},  
\end{eqnarray}
for $4<n<-3$ using a Gaussian window function. The question is: 
How the scaling law of perturbations, \sm, evolves starting
from an initial $(\sm)_i$?

In the early 1970s, Harrison and Zel'dovich independently asked themselves 
about the functionality of $\sm$ (or the density contrast) at the time adiabatic
perturbations cross the horizon, that is, if $(\sm)_H\propto M^{\alpha_H}$, then 
what is the value of $\alpha_H$? These authors concluded that $-0.1\le\alpha_H\le 0.2$,
i.e. $\alpha_H\approx 0$ ($n_H\approx -3$). If  $\alpha_H>>0$ ($n_H>>-3$), then
$\sm\rightarrow \infty$ for $M\rightarrow 0$; this means that for a given small 
mass scale $M$,
the mass density of the perturbation at the time of becoming causally connected
can correspond to the one of a (primordial) black hole. Hawking evaporation of 
black holes put a constraint on $M_{BH,prim}\lesssim 10^{15}$g, which corresponds
to $\alpha_H\le 0.2$, otherwise the $\gamma$--ray background radiation would be 
more intense than that observed. If  $\alpha_H<<0$ ($n_H<<-3$), then larger scales
would be denser than the small ones, contrary to what is observed. The 
scale--invariant {\it Harrison--Zel'dovich power spectrum, $P_H(k)\propto k^{-3}$,} 
is for perturbations
at the time of entering the horizon. How should the primordial power spectrum
$P_i(k)=Ak^n_i$ or $(\sm)_i=BM^{-\alpha_i}$ (defined at some fixed initial time) 
be to produce such scale invariance?  Since $t_i$ until the horizon crossing time 
$t_{cross}(M)$ for a given perturbation of mass $M$, $\sm(t)$ evolves as $a(t)^2$ 
(supra--horizon regime in the radiation era).
At $t_{cross}$,  the horizon mass $M_H$ is equal by definition to $M$. We have
seen that $M_H\propto a^3$ (radiation dominion), so that $\ac\propto M_H^{1/3}=M^{1/3}$.
Therefore, 
\begin{equation}
\sm(t_{cross})\propto (\sm)_i(\ac/a_i)^2\propto M^{-\alpha_i}M^{2/3},
\end{equation}
i.e. $\alpha_H=2/3-\alpha_i$  or $n_H=n_i-4$. A similar result is obtained if the 
perturbation enters the horizon during the matter dominion era. From this analysis
one concludes that for the perturbations to be scale invariant at horizon
crossing ($\alpha_H=0$ or $n_H=-3$), the primordial (initial) power spectrum 
should be $P_i(k)=Ak^1$ or $(\sm)_i\propto M^{-2/3}\propto \lambda_0^{-2}$ 
(i.e. $n_i=1$ and  $\alpha=2/3$; $A$ is a normalization constant). Does inflation 
predict such power spectrum?
We have seen that, according to the quantum field theory and assuming that
$H=$const during inflation,
the fluctuation amplitude is scale invariant at the time to 
exit the horizon, $\delta_H\sim$const. On the other hand, we have seen
that supra--horizon curvature perturbations during a de Sitter period
evolve as $\delta\propto a^{-2}$  (eq. 4). Therefore, at the end
of inflation we have that $\delta_{inf}=\delta_H(\lambda_0)(a_{inf}/a_H)^{-2}$.
The proper size of the fluctuation when crossing the horizon is
$\lambda_p=a_H\lambda_0\approx H^{-1}$; therefore, $a_H\approx 1/(\lambda_0 H)$.
Replacing now this expression in the equation for $\delta_{inf}$
we get that: 
\begin{equation}
\delta_{inf}\approx\delta_H(\lambda_0)(a_{inf}\lambda_0 H)^{-2}\propto
\lambda_0^{-2}\propto M^{-2/3},
\end{equation} 
if $\delta_H\sim$const. Thus, inflation  predicts $\alpha_i$ 
nearly equal to $2/3$ ($n_i\approx 1$)!  
Recent results from the analysis of CMBR anisotropies by the \textit{WMAP}
satellite \cite{spergel06} seem to show that $n_i$ is slightly smaller than 1 or 
that $n_i$ changes with the scale (running power--spectrum index). This
is in more agreement with several inflationary models, where $H$ actually
slightly vary with time introducing some scale dependence in $\delta_H$. 

\begin{figure}
\centering
\includegraphics[height=6.5cm]{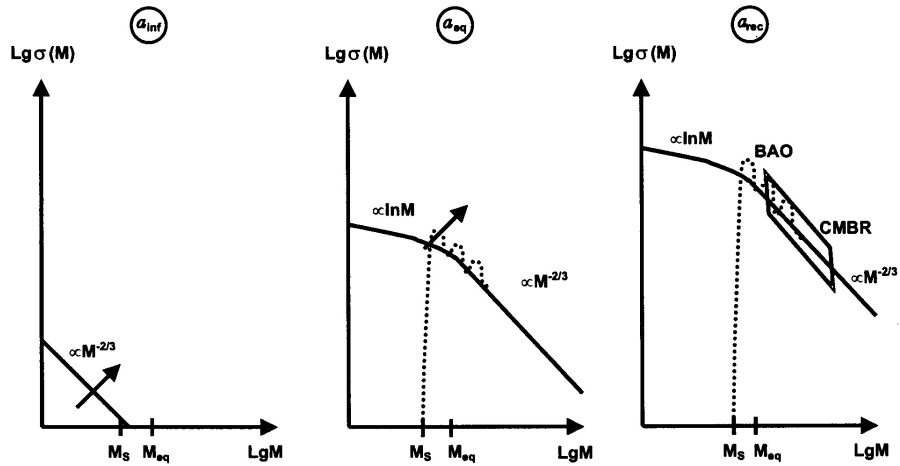}
\caption{Linear evolution of the perturbation mass variance \sm. The perturbation
amplitude in the supra--horizon regime grow kinematically. DM perturbations (solid 
curve) that cross the horizon during the radiation dominion, freeze--out their 
grow due to stangexpantion, producing a flattening in the scaling law \sm\ for all 
scales smaller than the corresponding to the horizon at the equality epoch 
(galaxy scales). Baryon--photon
perturbations smaller than the Silk mass $M_S$ are damped out (dotted curve) 
and those larger than $M_S$ but smaller than the horizon mass at recombination 
are oscillating (Baryonic Acoustic Oscillation, BAO).}
\label{fluctMO}       
\end{figure}

The perturbation analysis, whose bases were presented in \S 3.2 and
resumed in Fig. \ref{resumen}, show us that \sm\ grows (kinematically) while 
perturbations are in the  supra--horizon regime. Once perturbations enter the 
horizon (first the smaller ones), if they are made of CDM, then the gravitational 
growth is ``freezed out'' whilst radiation dominates (stangexpantion). As shown
schematically in Fig. \ref{fluctMO}, this ``flattens'' the variance \sm\ 
at scales smaller than $M_{H,eq}$; in fact, $\sm\propto ln(M)$ at these
scales, corresponding to galaxies! After \te\ the CDM variance
(or power spectrum) grows at the same rate at all scales. If perturbations
are made out of baryons, then for scales smaller than $M_S$, the gravito--acoustic
oscillations are damped out, while for scales close to the Hubble radius at 
recombination, these oscillations are present. The ``final'' processed 
mass variance or power spectrum is defined at the recombination epoch.
For example, the power spectrum is expressed as: 
\begin{equation}
P_{rec}(k)= Ak^{n_i}\times (D(\trec)/D(t_i))^2\times T^2(k), 
\end{equation}
where the first term is the initial power spectrum $P_i(k)$; the second one is how much
the fluctuation amplitude has grown in the linear regime ($D(t)$ is the 
so--called linear growth factor), and the third one is a transfer function
that encapsulates the different damping and freezing out processes able to 
deform the initial power spectrum shape. At large scales, $T^2(k)=1$, i.e.
the primordial shape is conserved (see Fig. \ref{fluctMO}).

Besides the mass power spectrum, it is possible to calculate the {\it angular power
spectrum of temperature fluctuation in the CMBR}.  This spectrum consists
basically of 2 ranges divided by a critical angular scales: 
the angle $\theta_{h}$ 
corresponding to the horizon scale at the epoch of recombination 
($(\lh)_{rec}\approx 200(\Omega h^2)^{-1/2}$ Mpc, comoving). For scales grander than 
$\theta_{h}$ the spectrum is featureless and corresponds to the scale--invariant
supra--horizon Sachs-Wolfe fluctuations. For scales smaller than $\theta_{h}$,
the sub--horizon fluctuations are dominated by the Doppler scattering 
(produced by the gravito--acoustic oscillations) with a series of decreasing
in amplitude peaks; the position (angle) of the first Doppler peak 
depends strongly on $\Omega$, i.e. on the geometry of the Universe. 
In the last 15 years, high--technology experiments as \textit{COBE}, 
\textit{Boomerang},  \textit{WMAP} provided valuable information (in particular
the latter one) on CMBR anisotropies. The results of this exciting branch of 
astronomy (called sometimes anisotronomy) were of paramount importance for astronomy 
and cosmology (see for a review \cite{hu02} and the W. Hu
website\footnote{\textit{http://background.uchicago.edu/$\sim$whu/physics/physics.html}}). 

Just to highlight some of the key results of CMBR studies, let us mention the 
next ones: 1) detailed predictions of the \Lcdm\ scenario 
concerning the linear evolution of perturbations were accurately proved, 2) 
several cosmological parameters as the geometry of the Universe, the baryonic 
fraction $\Omega_B$, and the index of the primordial power spectrum, were determined
with high precision (see the actualized, recently delivered results
from the 3 year analysis of \textit{WMAP} in \cite{spergel06}), 3) by studying
the polarization maps of the CMBR it was possible to infer the epoch when
the Universe started to be significantly reionized by the formation of 
first stars, 4) the amplitude (normalization) of the primordial fluctuation 
power spectrum was accurately measured. The latter 
is crucial for further calculating the non--linear regime of cosmic structure
formation. I should emphasize that while the shape of the power spectrum is 
predicted and well understood within the context of the \Lcdm\ model, the 
situation is fuzzy concerning the power spectrum normalization. We have a 
phenomenological value for $A$ but not a theoretical prediction.

\section{The dark side of galaxy formation and evolution}

A great triumph of the \Lcdm\ scenario was the overall consistency found
between predicted and observed CMBR anisotropies generated at the recombination
epoch. In this scenario, the gravitational evolution of CDM perturbations is 
the driver of cosmic structure formation. At scales much larger than galaxies, 
(i) mass density perturbations are still in the (quasi)linear regime, following
the scaling law of primordial fluctuations, and (ii) the dissipative physics of 
baryons does not affect significantly the matter distribution. Thus, the 
large--scale structure (LSS) of the Universe is determined 
basically by DM perturbations yet in their (quasi)linear regime. 
At smaller scales, non--linearity strongly affects the primordial scaling law
and, moreover, the dissipative physics of baryons ``distorts'' the
original DM distribution,  particularly inside galaxy--sized DM halos. 
However, DM in any case provides the original ``mold'' where gas dynamics
processes take place.

The \Lcdm\ scenario describes successfully the observed LSS of the Universe
(for reviews see e.g., \cite{frenk02,guzzo02}, and for some recent observational
results see e.g. \cite{tegmark04,seljak05,spergel06}).  The observed filamentary 
structure 
can be explained as a natural consequence of the CDM gravitational instability 
occurring preferentially in the shortest axis of 3D and 2D protostructures
(the Zel'dovich panckakes). 
The clustering of matter in space, traced mainly by galaxies, is also well 
explained by the clustering properties of CDM.  At scales $r$ much larger than 
typical galaxy sizes, the galaxy 2-point correlation function $\xi_{gal}(r)$ 
(a measure of the average clustering strength on spheres of radius $r$) agrees
rather well with $\xi_{CDM}(r)$. Current large statistical galaxy surveys as 
SDSS and 2dFGRS, allow now to measure the redshift--space 2-point 
correlation function at large scales with unprecedented accuracy, to the point 
that weak ``bumps'' associated with the baryon acoustic
oscillations at the recombination epoch begin to be detected \cite{Eisen05}.
At small scales ($\lesssim 3\mpch$), $\xi_{gal}(r)$ departs from the
predicted pure $\xi_{CDM}(r)$ due to the emergence of two processes:
(i) the strong non--linear evolution that small scales underwent, and (ii)
the complexity of the baryon processes related to galaxy formation. The
difference between  $\xi_{gal}(r)$ and $\xi_{CDM}(r)$ is parametrized
through one ``ignorance'' parameter, $b$, called bias, $\xi_{gal}(r)=b\xi_{CDM}(r)$.
Below, some basic ideas and results related to the former processes will
be described. The baryonic process will be sketched in the next Section.

\subsection{Nonlinear clustering evolution}

The scaling law of the processed \Lcdm\ perturbations, is such that
\sm\ at galaxy--halo scales decreases slightly with mass (logarithmically)
and for larger scales, decreases as a power law (see Fig. \ref{fluctMO}).
Because the perturbations of higher amplitudes collapse first, the 
first structures to form in the \Lcdm\ scenario are 
typically the smallest ones. Larger structures assemble from the smaller
ones in a process called {\it hierarchical clustering} or bottom--up
mass assembling. It is interesting to note that the concept of hierarchical
clustering was introduced several years before the CDM paradigm emerged.
Two seminal papers settled the basis for the current theory of galaxy
formation: Press \& Schechter 1974 \cite{PS74} and White \& Rees 1979 \cite{WR78}.
In the latter it was proposed that ``the  smaller--scale virialized [dark]
systems merge into an amorphous whole when they are incorporated in a larger 
bound cluster. Residual gas in the resulting potential wells cools and acquires 
sufficient concentration to self--gravitate, forming luminous galaxies up to a 
limiting size''. 

\begin{figure}
\centering
\includegraphics[height=7cm]{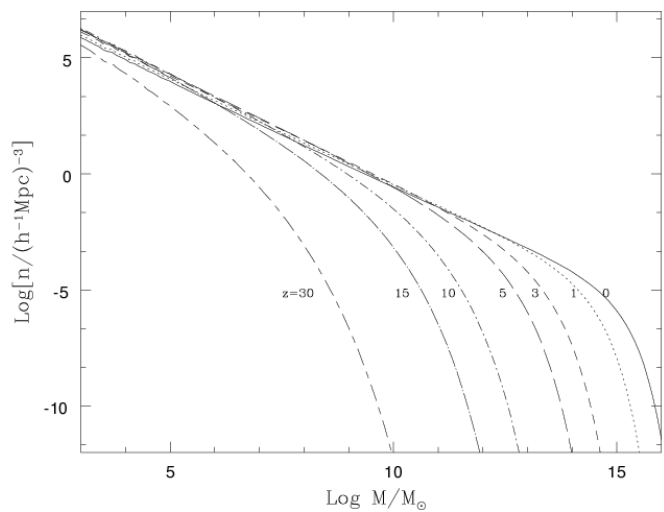}
\caption{Evolution of the comoving number density of collapsed halos
(P--S mass function) according to the ellipsoidal
modification by \cite{sheth}. Note that the ``cut--off'' mass grows
with time. Most of the mass fraction in collapsed
halos at a given epoch are contained in halos with masses around 
the ``cut--off'' mass.}
\label{PS}       
\end{figure}

The Press \& Schechter (P-S) formalism was developed to calculate the 
mass function (per unit of comoving volume) of halos at a given epoch,
$n(M,z)$. The starting point is a Gaussian density field filtered (smoothed)
at different scales corresponding to different masses, the mass variance \sm\ 
being the characterization of this filtering process.  A collapsed halo
is identified when the evolving density contrast of the region of mass $M$, 
$\delta_M(z)$, attains a critical value, $\delta_c$, given by the 
spherical top--hat collapse model\footnote{The spherical top--hat model refers to the 
exact calculation of the collapse of a uniform spherical density perturbation
in an otherwise uniform Universe; the dynamics of such a region is the same
of a closed Universe. The solution of the equations of motion shows that
the perturbation at the beginning expands as the background Universe (proportional
to $a$), then it reaches a maximum expansion (size) in a time $t_{max}$, and since
that moment the perturbation separates of the expanding background, collapsing
in a time $t_{col}=2t_{max}$. }. This way, the Gaussian probability 
distribution for $\delta_M$ is used to calculate the mass distribution of
objects collapsed at the epoch $z$. The P-S formalism assumes
implicitly that the only objects to be counted as collapsed halos at a given
epoch are those with $\delta_M(z)=\delta_c$. For a mass variance decreasing
with mass, as is the case for CDM models, this implies a ``hierarchical'' 
evolution of $n(M,z)$: as $z$ decreases, less massive collapsed
objects disappear in favor of more massive ones (see Fig. \ref{PS}). 
The original P-S formalism had an error of 2 in the sense that integrating 
$n(M,z)$ half of the mass is lost. The authors multiplied $n(M,z)$ by 2,
argumenting that the objects duplicate their masses by accretion from the 
sub--dense regions. The problem of the factor of 2 in the P-S analysis
was partially solved using an excursion set statistical approach \cite{Bond90,LC93}.

\begin{figure}
\centering
\includegraphics[height=5cm]{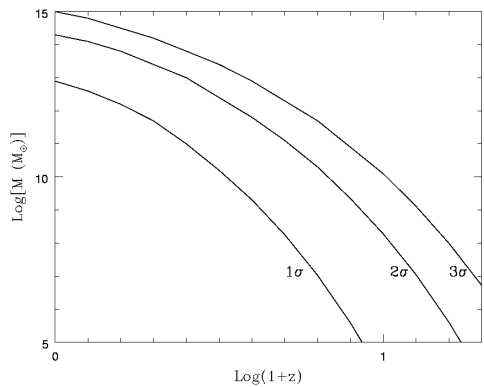}
%
%
\caption{Collapse redshifts of spherical top--hat 1$\sigma$, 2$\sigma$
and 3$\sigma$ perturbations in a \Lcdm\ cosmology with $\sigma_8=0.9$. 
Note that galaxy--sized ($M\sim 10^8-10^{13}\msun$) 1$\sigma$ halos collapse
in a redshift range, from $z\sim 3.5$ to $z=0$,
respectively; the corresponding ages are from $\sim 1.9$ to 13.8 Gyr, respectively.}
\label{colapsetime}       
\end{figure}

To get an idea of the typical formation epochs of CDM halos, the spherical
collapse model can be used. According to this model, the density contrast
of given overdense region, $\delta$, grows with $z$ proportional to the 
growing factor, $D(z)$, until it reaches a critical
value, $\delta _{c}$, after which the perturbation is supposed to collapse
and virialize\footnote{The mathematical solution gives that the spherical
perturbed region collapses into a point (a black hole) after reaching its
maximum expansion. However, real 
perturbations are lumpy and the particle orbits are not perfectly radial. 
In this situation, during the collapse the structure comes to a dynamical
equilibrium under the influence of large scale gravitational potential
gradients, a process named by the oxymoron ``violent relaxation'' (see
e.g. \cite{BT87}); 
this is a typical collective phenomenon. The end result is a system that
satisfies the virial theorem: for a self--gravitating system this means
that the internal kinetic energy is half the (negative) gravitational
potential energy. Gravity is supported by the velocity dispersion of particles
or lumps. The collapse factor is roughly 1/2, i.e. the typical
virial radius $R_v$ of the collapsed structure is $\approx 0.5$ the radius
of the perturbation at its maximum expansion.}.
at redshift $z_{\rm col}$ (for example see \cite{padma}): 
\begin{equation}
\delta (z_{\mathrm{col}})\equiv \delta _{0}D(z_{\mathrm{col}})=\delta
_{c,0}.  \label{delta}
\end{equation}%
The convention is to fix all the quantities to their linearly extrapolated
values at the present epoch (indicated by the subscript ``0'') in such a way
that $D(z=0)\equiv D_0=1$. Within this convention, for an Einstein--de Sitter
cosmology, $\delta _{c,0}=1.686$, while for the \Lcdm\ cosmology,
$\delta _{c,0}=1.686\Omega _{M,0}^{0.0055}$, and the growing factor is given by 
\begin{equation}
D(z)=\frac{g(z)}{g(z_{0})(1+z)},  \label{grow}
\end{equation}%
where a good approximation for $g(z)$ is \cite{carroll}: 
\begin{equation}
g(z)\simeq \frac{5}{2}\left[ \Omega _{M}^{\frac{4}{7}}-\Omega _{\Lambda
}+\left( 1+\frac{\Omega _{M}}{2}\right) \left( 1+\frac{\Omega _{\Lambda }}{70%
}\right) \right] ^{-1},  \label{gz}
\end{equation}%
and where $\Omega _{M}=\Omega _{M,0}(1+z)^{3}/E^{2}(z)$, $\Omega _{\Lambda
}=\Omega _{\Lambda }/E^{2}(z)$, with $E^{2}(z)=\Omega _{\Lambda }+\Omega
_{M,0}\left( 1+z\right) ^{3}$. For the Einstein--de Sitter model, $D(z)=(1+z)$.
We need now to connect the top--hat sphere results to a perturbation of mass $M$. 
The processed perturbation field, fixed at the present epoch, is characterized 
by the mass variance \sm\ and we may assume that $\delta_0=\nu\sm$, where 
$\delta_0$ is $\delta$ linearly extrapolated to $z=0$, and  $\nu$ is the peak 
height. For average perturbations, $\nu=1$, while for rare, high--density 
perturbations, from which the first structures arose, $\nu>>1$.  By introducing 
 $\delta_0=\nu\sm$ into eq. (\ref{delta}) one may infer $z_{col}$ for
a given mass. Fig. \ref{colapsetime} shows the typical $z_{col}$ 
of $1\sigma$,  $2\sigma$, and  $3\sigma$ halos.  The collapse
of galaxy--sized $1\sigma$ halos occurs within a relatively small range
of redshifts. This is a direct consequence of the ``flattening'' suffered
by \sm\ during radiation--dominated era due to stangexpansion (see \S 3.2).
Therefore, in a \Lcdm\ Universe it is not expected to observe a significant
population of galaxies at $z\grtsim 5$. 

%

The problem of cosmological gravitational clustering is very complex due to 
non--linearity, lack of symmetry and large dynamical range. Analytical and
semi--analytical approaches provide illuminating results but numerical 
N--body simulations are necessary to tackle all the aspects of this problem.
In the last 20 years, the ``industry'' of numerical simulations had an 
impressive development. The first cosmological simulations in the middle 80s
used a few $10^4$ particles (e.g., \cite{Davis85}). The currently largest simulation
(called the Millenium simulation \cite{springel05}) uses $\sim 10^{10}$ particles! 
A main effort  is done to reach larger and larger dynamic ranges in order to simulate 
encompassing volumes large enough to contain representative populations of 
all kinds of halos (low mass and massive ones, in low-- and high--density 
environments,  high--peak rare halos),  yet resolving the inner structure of 
individual halos.

\paragraph{Halo mass function}

The CDM halo mass function (comoving number density of halos of different masses
at a given epoch $z$, $n(M,z)$) obtained in the N--body simulations is consistent
with the P-S function in general, which is amazing given the approximate
character of the P-S analysis. However, in more detail, the results of large
N--body simulations are better fitted by modified P-S analytical functions, 
as the one derived in \cite{sheth} and showed in Fig. \ref{PS}. Using the
Millennium simulation, the halo mass function has been accurately measured
in the range that is well sampled by this run 
($z\le 12, M\ge 1.7\times 10^{10}\msunh$). The mass function is described
by a power law at low masses and an exponential cut--off at larger masses.
The ``cut-off'', most typical mass, increases with time and is
related to the hierarchical evolution of the $1\sigma$ halos shown in
Fig. \ref{colapsetime}. The halo mass function is the starting point for 
modeling the luminosity function of galaxies. From Fig. \ref{PS} we
see that the evolution of the abundances of massive halos is much more 
pronounced than the evolution of less massive halos. This is why 
observational studies of abundance of massive galaxies or cluster
of galaxies at high redshifts provide a sharp test to theories
of cosmic structure formation. The abundance of massive rare halos
at high redshifts are for example a strong function of the fluctuation
field primordial statistics (Gaussianity  or non-Gaussianity).

\subparagraph{Subhalos.}
An important result of N--body simulations is the existence of 
subhalos, i.e. halos inside the virial radius of larger halos, which 
survived as self--bound entities the gravitational collapse of the higher 
level of the hierarchy. 
Of course, subhalos suffer strong mass loss due to tidal stripping,
but this is probably not relevant for the luminous galaxies formed
in the innermost regions of (sub)halos. This is why in the case of subhalos, 
the maximum circular velocity $V_m$ (attained at radii much smaller than
the virial radius) is used instead of the virial mass. The $V_m$ distribution 
of subhalos inside cluster--sized and galaxy--sized halos is similar \cite{moore}.
This distribution agrees with the distribution of galaxies seen in clusters,
but for galaxy--sized halos the number of subhalos overwhelms by 1--2
orders of magnitude the observed number of satellite galaxies around galaxies
like Milky Way and Andromeda \cite{klypin,moore}. 

\begin{figure}
\centering
\includegraphics[height=8.7cm]{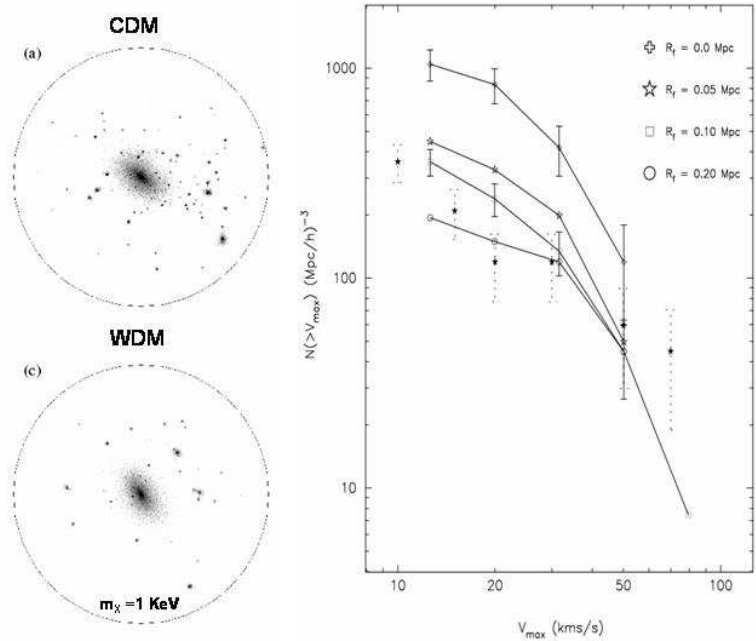}
\caption{Dark matter distribution in a sphere of 400\mpch\ of a simulated
Galaxy--sized halo with CDM (a) and WDM ($m_X=1$KeV, b). The substructure
in the latter case is significantly erased. Right panel shows the cumulative
maximum $V_c$ distribution for both cases (open crosses and squares, respectively)
as well as for an average of observations of satellite galaxies in our Galaxy 
and in Andromeda (dotted error bars). \textit{Adapted from \cite{pedro}.} }
\label{pedro}       
\end{figure}

Fig. \ref{pedro} (right side) shows the subhalo cumulative $V_m-$distribution 
for a CDM Milky Way--like halo compared to the
observed satellite $V_m-$distribution. In this Fig. are also shown
the $V_m-$distributions obtained for the same Milky--Way halo 
but using the power spectrum of three WDM models with particle masses
$m_X\approx 0.6, 1$, and 1.7 KeV. The smaller $m_X$, the larger is the
free--streaming (filtering) scale, $R_f$, and the more substructure is washed 
out (see \S 3.2).
In the left side of Fig. \ref{pedro} is shown the DM distribution
inside the Milky--Way halo simulated by using a CDM power spectrum (top)
and a WDM power spectrum with $m_X\approx 1$KeV (sterile neutrino, bottom).
For a student it should be exciting to see with her(his) own eyes 
this tight connection between micro-- and macro--cosmos: the mass of 
the elemental particle determines the structure and substructure
properties of galaxy halos!

\paragraph{Halo density profiles}

High--resolution N--body simulations \cite{NFW} and semi--analytical
techniques (e.g., \cite{AFH}) allowed to answer the following questions: How is the
inner mass distribution in CDM halos? Does this distribution depend
on mass? How universal is it?  The two--parameter density profile 
established in \cite{NFW} (the Navarro-Frenk-White, NFW profile) departs 
from a single power law, and it was proposed to be universal and not 
depending on mass. In fact the slope $\beta(r)\equiv -$dlog$\rho(r)/$dlog$r$ of the 
NFW profile changes from $-1$ in the center to $-3$ in the periphery. The
two parameters, a normalization factor, $\rho_s$ and a shape factor,
$r_s$, were found to be related in a such a way that the profile
depends only on one shape parameter that could be expressed as the concentration,
$c_{NFW}\equiv r_s/R_v$. The more massive the halo, the less concentrated
on the average. For the \Lcdm\ model, $c\approx 20-5$ for 
$M\sim 2\times 10^8-2\times 10^{15}\msunh$, respectively \cite{eke}. However,
for a given $M$, the scatter of $c_{NFW}$ is large ($\approx 30-40\%$),
and it is related to the halo formation history \cite{AFH,bullock01a,wechsler}
(see below). A significant fraction of halos depart from the NFW profile.
These are typically not relaxed or disturbed by companions or external
tidal forces.

\subparagraph{Is there a ``cusp'' crisis?}

More recently, it was found that the inner 
density profile of halos can be steeper than $\beta=-1$ (e.g. \cite{moore99}).
However, it was shown that in the limit of resolution, $\beta$ never
is as steep a $-1.5$ \cite{navarro04}.  The inner structure
of CDM halos can  be tested in principle with observations of (i) the inner rotation
curves of DM dominated galaxies (Irr dwarf and LSB galaxies; the inner velocity 
dispersion of dSph galaxies is also being used as a test ), and (ii)
strong gravitational lensing and hot gas distribution in the inner regions of 
clusters of galaxies. Observations suggest that the DM distribution in dwarf
and LSB galaxies has a roughly constant density core, in contrast to the cuspy 
cores of CDM halos (the literature on this subject is extensive; see for
recent results \cite{deblok,gentile04,simon04,weldrake03} and more references therein). 
If the observational studies confirm that halos have constant--density cores, then
either astrophysical mechanisms able to expand the halo cores should work efficiently
or the \Lcdm\ scenario should be modified. In the latter
case, one of the possibilities is to introduce weakly self--interacting
DM particles. For small cross sections, the interaction is effective
only in the more dense inner regions of galaxies, where heat inflow
may expand the core. However, the gravo--thermal catastrophe can also be 
triggered. In \cite{colin} it was shown that in order to avoid the 
gravo--thermal instability
and to produce shallow cores with densities approximately constant
for all masses, as suggested by observations, the DM cross section per unit
of particle mass should be $\sigma_{DM}/m_X=0.5-1.0v_{100}^{-1}$ cm$^2$/gr,
where $v_{100}$ is the relative velocity of the colliding particles in 
unities of 100 km/s; $v_{100}$ is close to the halo maximum circular
velocity, $V_m$.  
 
The DM mass distribution was inferred from the rotation curves of dwarf and
LSB galaxies under the assumptions of circular motion, halo spherical symmetry,
the lack of asymmetrical drift, etc. In recent studies it was discussed
that these assumptions work typically in the sense of lowering the observed
inner rotation velocity \cite{hayashi04,rhee04,valenzuela}. For example, in 
\cite{valenzuela} it is demonstrated
that non-circular motions (due to a bar) combined with gas pressure support 
and projection effects systematically underestimate by up to 50\% the 
rotation velocity of cold gas in the central 1 kpc region of their 
simulated dwarf galaxies, creating the illusion of a constant density core.

\subparagraph{Mass--velocity relation.} In a very simplistic analysis, it
is easy to find that $M\propto V_c^3$ if the average halo density 
$\rho_h$ does not depend on mass. On one hand, $V_c\propto (GM/R)^{1/2}$,
and on the other hand, $\rho_h\propto M/R^3$, so that 
$V_c\propto M^{1/3}\rho_h^{1/6}$. Therefore, for $\rho_h=$const, $M\propto V_c^3$.
We have seen in \S 3.2 that the CDM perturbations at galaxy scales have similar 
amplitudes (actually $\sm\propto$ ln$M$) due to the stangexpansion effect in the 
radiation--dominated era. This implies that galaxy--sized perturbations
collapse within a small range of epochs attaining more or less similar average
densities. The CDM halos actually have a mass distribution that translates
into a circular velocity profile $V_c(r)$. The maximum of this profile, $V_m$,
is typically the circular velocity that characterizes a given halo of virial
mass $M$. Numerical and semi--numerical results show that (\Lcdm\ model): 
\begin{equation}
M\approx 5.2\times 10^4\left(\frac{V_m}{km s^{-1}}\right)^{3.2}\msunh,
\end{equation}
Assuming that the disk infrared luminosity $L_{IR}\propto M$, and that the disk 
maximum rotation velocity $V_{rot,m}\propto V_m$, one obtains that 
$L_{IR}\propto V_{rot,m}^{3.2}$, amazingly similar to the observed
infrared Tully--Fisher relation \cite{tully}, one of the most robust and 
intriguingly correlations in the galaxy world! I conclude that this
relation is a {\it clear imprint of the CDM power spectrum of fluctuations.}

\paragraph{Mass assembling histories}

\begin{figure}
\centering
\includegraphics[height=9.4cm]{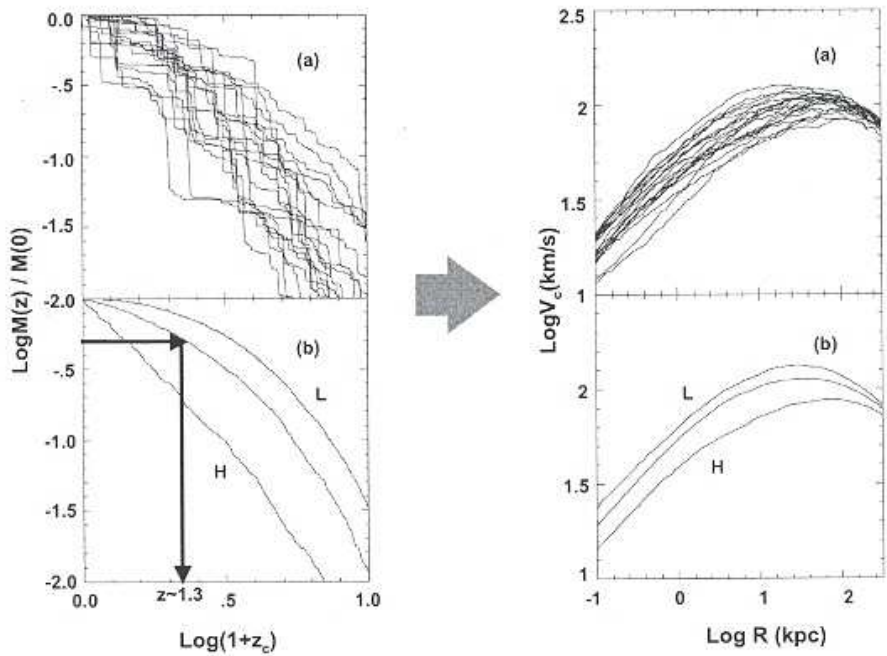}
\caption{\textit{Upper panels (a).} A score of random halo MAHs for a 
present--day virial mass of $3.5\times 10^{11}\msun$ and the corresponding 
circular velocity profiles of the virialized halos. \textit{Lower panels (b).} 
The average MAH and two extreme deviations from $10^4$ random MAHs for
the same mass as in (a), and the corresponding halo circular
velocity profiles. The MAHs are diverse for a given mass and the $V_c$ (mass)
distribution of the halos depend on the MAH. \textit{Adapted from \cite{FA00}.} }
\label{MAHs}       
\end{figure}

One of the key concepts of the hierarchical clustering scenario is
that cosmic structures form by a process of continuous mass aggregation,
opposite to the monolithic collapse scenario. The mass assembly of CDM halos 
is characterized by the mass aggregation history (MAH), which can alternate
{\it smooth mass accretion} with {\it violent major mergers}. The MAH can be 
calculated by using semi--analytical approaches
based on extensions of the P-S formalism. The main idea lies in 
the estimate of the {\it conditional} probability that given
a collapsed region of mass $M_0$ at $z_0$, a region of mass $M_1$  
embedded within the volume containing $M_0$, had collapsed at an 
earlier epoch $z_1$. This probability is calculated based on the
excursion set formalism starting from a Gaussian density field
characterized by an evolving mass variance \sm \cite{Bond90,LC93}. 
By using the conditional probability and random trials at each temporal
step, the ``backward'' MAHs corresponding to a fixed mass $M_0$ (defined 
for instance at $z=0$) can be traced. The MAHs of isolated halos by definition
decrease toward the past, following different tracks (Fig. \ref{MAHs}), 
sometimes with abrupt big jumps that can be identified as major mergers 
in the halo assembly history. 

To characterize typical behaviors of the halo MAHs, one may calculate
the average MAH for a given virial mass $M_0$, for a given ``population'' of
halos selected by its environment, etc. In the left panels of Fig. \ref{MAHs} 
are shown 20 individual MAHs randomly selected from 10$^4$ trials for 
$M_0=3.5\times 10^{11}\msun$ in a \Lcdm\ cosmology \cite{FA00}. In the
bottom panel are plotted the average MAH from these 10$^4$ trials as
well as two extreme deviations from the average. The average MAHs
depend on mass: more massive halos have a more extended average MAH,
i.e. they aggregate a given fraction of $M_0$ latter than less massive
halos. It is a convention to define the typical halo formation redshift,
$z_f$, when half of the current halo mass $M_0$ has been aggregated. For 
instance, for the \Lcdm\ cosmology the average MAHs show that $z_f\approx 2.2, 
1.2$ and 0.7 for $M_0=10^{10}\msun, 10^{12}\msun$ and $10^{14}\msun$, respectively.
A more physical definition of halo formation time is when the halo maximum
circular velocity $V_m$ attains its maximum value. After this epoch,
the mass can continue growing, but the inner gravitational potential of the
system is already set. 

Right panels of Fig. \ref{MAHs} show the present--day halo circular velocity
profiles, $V_c(r)$, corresponding to the MAHs plotted in the left panels. 
The average $V_c(r)$ is well described by the NFW profile. There is a direct
relation between the MAH and the halo structure as described by $V_c(r)$ 
or the concentration parameter. The later the MAH, the more extended
is $V_c(r)$ and the less concentrated is the halo \cite{AFH,wechsler}. 
Using high--resolution simulations some authors have shown that
the halo MAH presents two regimes: an early phase of fast mass 
aggregation (mainly by major mergers) and a late phase of slow aggregation 
(mainly by smooth mass accretion) \cite{zhao03a,Lin05}. 
The potential well of a present--day halo is set mainly at the end of the 
fast, major--merging driven, growth phase.   

From the MAHs we may infer: (i) the mass aggregation rate evolution of
halos (halo mass aggregated per unit of time at different $z'$s), and (ii) the
major merging rates of halos (number of major mergers per unit of time
per halo at different $z'$s). These quantities should be closely related to the
star formation rates of the galaxies formed within the halos as well as
to the merging of luminous galaxies and pair galaxy statistics. By using
the \Lcdm\ model, several studies showed that most of the mass of the
present--day halos has been aggregated by accretion rather than major 
mergers (e.g., \cite{murali}). Major merging was more frequent in the past 
\cite{gott01}, and it 
is important for understanding the formation of massive galaxy spheroids and the 
phenomena related to this process like QSOs, supermassive black hole
growth, obscured star formation bursts, etc. Both the mass aggregation rate
and major merging rate histories depend strongly on environment: the denser
the environment, the higher is the merging rate in the past. However,
in the dense environments (group and clusters) form typically structures more
massive than in the less dense regions (field and voids). Once a large
structure virializes, the smaller, galaxy--sized halos become subhalos
with high velocity dispersions: the mass growth of the subhalos is truncated,
or even reversed due to tidal stripping, and the merging probability 
strongly decreases. Halo assembling (and therefore, galaxy assembling)
definitively depends on environment. Overall, by integrating the MAHs
of the whole galaxy--sized \Lcdm\ halo population in a given volume, the 
general result is that the peak in halo assembling activity was at $z\approx 1-2$.
After these redshifts, the global mass aggregation rate strongly decreases
(e.g., \cite{vdB02}. 

To illustrate the driving role of DM processes in galaxy evolution, I 
mention briefly here two concrete examples: 

{\it 1). Distributions of present--day
specific mass aggregation rate, $(\dot{M}/M)_0$,  and halo lookback 
formation time, $T_{1/2}$.} For a \Lcdm\ model, these distributions
are bimodal, in particular the former. We have found that roughly $40\%$ of halos 
(masses larger than $\approx 10^{11}\msunh$) have $(\dot{M}/M)_0\le 0$;
they are basically subhalos. The remaining 60\% present a broad 
distribution of $(\dot{M}/M)_0>0$ peaked at $\approx 0.04$Gyr$^{-1}$.
Moreover, this bimodality strongly changes with large--scale environment:
the denser is the environment the, higher is the fraction of halos
with $(\dot{M}/M)_0\le 0$. It is interesting enough that similar fractions
and dependences on environment are found for the specific star formation
rates of galaxies in large statistical surveys (\S\S 2.3); the situation
is similar when confronting the distributions of $T_{1/2}$ and observed
colors. Therefore, it seems that the {\it the main driver of the observed
bimodalities in $z=0$ specific star formation rate and color of galaxies is
the nature of the CDM halo mass aggregation process.} Astrophysical processes
of course are important but the main body of the bimodalities can be
explained just at the level of DM processes. 
  
{\it 2. Major merging rates.} The observational inference of galaxy major 
merging rates is not an easy task. The two commonly used methods are based on
the statistics of galaxy pairs (pre--mergers) and in the morphological 
distortions of ellipticals (post--mergers). The results show that the 
merging rate increases as $(1+z)^x$, with $x\sim 0-4$. The predicted 
major merging rates in the \Lcdm\ scenario agree roughly with 
those inferred from statistics of galaxy pairs. From the fraction 
of normal galaxies in close companions (with separations less than 
50 kpch$^{-1}$) inferred from observations
at $z=0$ and $z=0.3$  \cite{Patton02}, and assuming an average 
merging time of $\sim 1$ Gyr for these separations, we estimate that 
the major merging rate at the present epoch is $\sim 0.01$ Gyr$^{-1}$ for 
halos in the range of $0.1-2.0 \ 10^{12}\msun$, while at $z=0.3$ the rate 
increased to $\sim 0.018$ Gyr$^{-1}$. These values are only slightly 
lower than predictions for the \Lcdm\ model.

\paragraph{Angular momentum}

The origin of the angular momentum (AM) is a key ingredient in theories of 
galaxy formation. Two mechanisms of AM acquirement were proposed for the
CDM halos (e.g., \cite{Peebles,Bullock01,maller02}): 1. tidal torques of 
the surrounding shear field when the 
perturbation is still in the linear regime, and 2. transfer of orbital
AM to internal AM in major and minor mergers of collapsed halos. The angular
momentum of DM halos is parametrized in terms of the dimensionless
spin parameter $\lambda\equiv J\sqrt{E}/(GM^{5/2}$, where $J$ is the modulus
of the total angular momentum and $E$ is the total (kinetic plus potential).
It is easy to show that $\lambda$ can be interpreted as the level of 
rotational support of a gravitational system, $\lambda= \omega/\omega_{sup}$,
where $\omega$ is the angular velocity of the system and $\omega_{sup}$
is the angular velocity needed for the system to be rotationally supported
against gravity (see \cite{padma}). 

For disk and elliptical galaxies, 
$\lambda\sim 0.4-0.8$ and $\sim 0.01-0.05$, respectively. Cosmological N--body 
simulations showed that the CDM halo spin parameter is log--normal distributed,
with a median value $\lambda\approx 0.04$ and a standard deviation
$\sigma_{\lambda}\approx 0.5$; this distribution is almost independent
from cosmology. A related quantity, but more
straightforward to compute is $\lambda'\equiv \frac{J}{\sqrt{2} M V_v R_v}$
\cite{Bullock01}, where $R_v$ is the virial radius and $V_v$ the
circular velocity at this radius. Recent simulations show that 
$(\lambda',\sigma_{\lambda'})\approx(0.035,0.6)$, though some variations
with environment and mass are measured \cite{AR05}. The evolution of 
the spin parameter depends on the AM acquirement mechanism. In general,
a significant systematical change of $\lambda$
with time is not expected, but relatively strong changes are measured in short
time steps, mainly after merging of halos, when $\lambda$ increases.

How is the internal AM distribution in CDM halos? Bullock et al. \cite{Bullock01}
found that in most of cases this distribution can be described by a simple
(universal) two--parameter function that departs significantly from the 
solid--body rotation distribution. In addition, the spatial distribution of 
AM in CDM halos tends to be cylindrical, being well aligned for 80\% of the halos,
and misaligned at different levels for the rest. The mass distribution
of the galaxies formed within CDM halos, under the assumption of specific
AM conservation, is established by $\lambda$, the halo AM distribution, and 
its alignment.

\subsection{Non--baryonic dark matter candidates}

The non--baryonic DM required in cosmology to explain observations and 
cosmic structure formation should be in form of elemental or scalar field 
particles or early formed quark nuggets. Modifications to fundamental physical
theories (modified Newtonian Dynamics, extra--dimensions, etc.)
are also plausible if DM is not discovered.

There are several docens of predicted elemental particles as DM candidates. 
The list is reduced if we focus only on well--motivated
exotic particles from the point of view of particle physics theory alone
(see for a recent review \cite{gondolo}).
The most popular particles beyond the standard model are the {\it supersymmetric
(SUSY)} particles in supersymmetric extensions of the Standard Model of particle 
physics. Supersymmetry is a new symmetry of space--time introduced in the process 
of unifying the fundamental forces of nature (including gravity). An excellent
CDM candidate is the lightest stable SUSY particle under the requirement that 
superpartners are only produced or destroyed in pairs (called R-parity conservation).
This particle called {\it neutralino} is weakly interacting and massive (WIMP).
Other SUSY particles are the gravitino and the sneutrino; they are of WDM type. 
The predicted masses 
for neutralino range from $\sim 30$ to 5000 GeV.   
The cosmological density of neutralino (and of other thermal WIMPs) is naturally as
required when their interaction cross section is of the order of a weak cross 
section. The latter gives the possibility to detect neutralinos in laboratory.

The possible discovery of WIMPs relies on two main techniques: 

\subparagraph{(i) Direct detections.} The WIMP interactions with nuclei 
(elastic scattering) in ultra--low--background terrestrial targets may deposit 
a tiny amount of energy  
($<50$ keV) in the target material; this kinetic energy of the recoiling nucleus 
is converted partly into scintillation light or ionization energy and partly into 
thermal energy. Dozens of experiments worldwide -of cryogenic or scintillator type, 
placed in mines or underground laboratories, attempt to measure these energies. 
Predicted event rates for neutralinos range from 10$^{-6}$ to 
10 events per kilogram detector material and day. The nuclear recoil spectrum is 
featureless, but depends on the WIMP and target nucleus mass. To convincingly 
detect a WIMP signal, a specific signature from the galactic halo particles is 
important. The Earth's motion through the galaxy induces both a seasonal variation 
of the total event rate and a forward--backward asymmetry in a directional signal.
The detection of structures in the dark velocity space, as those predicted  
to be produced by the Sagittarius stream, is also an specific signature from the 
Galactic halo; directional detectors are needed to measure this kind of signatures.
 
The DAMA collaboration reported a possible detection of WIMP particles obeying
the seasonal variation; the most probable value of the WIMP mass was $\sim 60$ GeV.
However, the interpretation of the detected signal as WIMP particles is controversial. 
The sensitivity of current experiments (e.g., CDMS and EDEL-WEISS) limit already 
the WIMP--proton spin--independent cross sections to values 
$\lesssim 2 \ 10^{-42}-10^{-40}$cm$^{-2}$ for the range
of masses $\sim 50-10^4$ GeV, respectively; for smaller masses, the
cross--section sensitivities are larger, and WIMP signals were not detected. Future
experiments will be able to test the regions in the cross-section--WIMP mass diagram,
where most of models make certain predictions.

\subparagraph{(ii) Indirect detections.} We can search for WIMPS by looking for the 
products of their annihilation.  The flux of annihilation products is proportional 
to the square of the WIMP density, thus regions of interest are those where the WIMP
concentration is relatively high. There are three types of searches according 
to the place where WIMP annihilation occur: (i) in the Sun or the Earth, which gives 
rise to a signal in high-energy neutrinos; (ii) in the galactic halo, or in the halo 
of external galaxies, which generates $\gamma-$rays and other cosmic rays such as 
positrons and antiprotons; (iii) around black holes, specially around the black 
hole at the Galactic Center. The predicted radiation fluxes depend on the particle 
physics model used to predict the WIMP candidate and on astrophysical quantities 
such as the dark matter halo structure, the presence of sub--structure, and the 
galactic cosmic ray diffusion model.

 Most of WIMPS were in thermal equilibrium in the early Universe (thermal relics). 
Particles which were produced by a non-thermal mechanism and that never had the 
chance of reaching thermal equilibrium are called non-thermal 
relics (e.g., axions, solitons produced in phase transitions, WIMPZILLAs produced 
gravitationally at the end of inflation). From the side of WDM, the most
popular candidate are the $\sim 1$KeV sterile neutrinos. A sterile neutrino
is a fermion that has no standard model interactions other than a coupling to 
the standard neutrinos through their mass generation mechanism.  Cosmological
probes, mainly the power spectrum of Ly$\alpha$ forest at high redshifts,
constrain the mass of the sterile neutrino to values larger than $\sim 2$KeV.

\section{The bright side of galaxy formation and evolution}

The \Lcdm\ scenario of cosmic structure formation has been 
well tested for perturbations that are still in the linear or quasilinear
phase of evolution. These tests are based, among other cosmological probes,
on accurate measurements of:

$\bullet$ the CMBR temperature fluctuations at large and small angular scales 

$\bullet$ the large--scale mass power spectrum as traced by the spatial distribution of 
galaxies and cluster of galaxies, by the Ly$\alpha$ forest clouds, by maps of 
gravitational weak and strong lensing, etc. 

$\bullet$ the peculiar large--scale motions of galaxies\footnote{Recall that linear
theory relates the peculiar velocity, that is the velocity deviation from 
the Hubble flow, to the density contrast. It is said that the cosmological
velocity field is {\it potential}; any primordial rotational motion able
to give rise to a density perturbation decays
as the Universe expands due to angular momentum conservation.}.   

$\bullet$ the statistics of strong gravitational lensing (multiple--lensed arcs).

Although these cosmological probes are based on observations of luminous (baryonic)
objects, the physics of baryons plays a minor or indirect role in the properties of 
the linear mass perturbations. The situation is different at small (galaxy) 
scales, where perturbations went into the non--linear regime and the dissipative 
physics of baryons becomes relevant. The interplay of DM and baryonic processes
is crucial for understanding galaxy formation and evolution. The progress in this
field was mostly heuristic; the \Lcdm\ scenario provides the initial and
boundary conditions for modeling galaxy evolution, but the complex physics
of the baryonic processes, in the absence of fundamental theories, 
requires a model adjustment through confrontation with the observations.

Following, I will outline some key concepts, ingredients, and results of the galaxy
evolution study based on the \Lcdm\ scenario. Some of the 
pioneer papers in this field are those of Gunn \cite{gunn77}, White \& Reese 
\cite{WR78}, Fall \& Efstathiou \cite{FE80}, Blumental et al. \cite{blum84},
Davis et al. \cite{Davis85}, Katz \& Gunn \cite{KG91}, White \& Frenk \cite{WF91}, 
Kauffmann et al. \cite{kauff93}. For useful lecture notes and recent reviews see e.g., 
Longair \cite{Longair89,Longairbook}, White \cite{W96}, Steinmetz \cite{steinmetz96}, 
Firmani \& Avila-Reese \cite{FA03}.

The main methods of studying galaxy formation and evolution in the \Lcdm\
context are:

$\bullet$ Semi-analytical Models (e.g., \cite{WF91,kauff93,cole94,baugh96,SP99,cole00,
benson03,baugh05}), where the halo mass assembling histories
are calculated with the extended Press--Schechter formalism and galaxies
are seeded within the halos by means of phenomenological recipes. This method is
very useful for producing whole populations of galaxies at a given epoch and predicting
statistical properties as the luminosity function and the morphological mix.

$\bullet$ Semi-numerical Models (e.g, \cite{FA00,AF00,vdB00,boissier}), 
where the internal physics of the galaxies, including those of the halos, are 
modeled numerically but under simplifying assumptions; the initial and boundary 
conditions are
taken from the \Lcdm\ scenario by using the extended Press--Schechter 
formalism and halo AM distributions from simulations. This method is
useful to predict the local properties of galaxies and correlations 
among the global properties, as well as to follow the overall evolution
of individual galaxies. 

 $\bullet$ Numerical N--body+hydrodyamical simulations (e.g., \cite{KG91,cen92,
katz92,navarro93,steinmetz94,weil98,abadi03,springel03,governato04}), where the 
DM and baryonic processes are followed in cosmological simulations. This is the 
most advanced and complete approach to galaxy evolution. However, current limitations
in the computational capabilities and the lack of fundamental theories 
for several of the physical processes involved, do not allow yet
to exploit optimally this method.  A great advance is being made
currently with an hybrid approach: in the high--resolution cosmological
N--body simulations of only DM, galaxies are grafted by using the 
semi--analytical models (e.g., \cite{Kaufetal99,
Helly03,deLucia04b,Berlind05,springel05,Kang05}).

\subsection{Disks}

The formation of galaxy disks deep inside the CDM halos is a generic process in 
the \Lcdm\ scenario. Let us outline the (simplified) steps of disk galaxy formation 
in this scenario:

\subparagraph{1. DM halo growth.} The ``mold''  for disk formation is 
provided by the mass and AM 
distributions of the virialized halo, which grows hierarchically. A description
of these aspects were presented in the previous Section.  
     
\subparagraph{2. Gas cooling and infall, and the maximum mass of galaxies.}  
It is common to assume that the gas in a halo is shock--heated during 
collapse to the virial temperature  \cite{WR78}. The gas then cools
radiatively and falls in a free--fall time to the center. The cooling function
$\Lambda(n,T_k;Z)$ depends on the gas density, temperature, and composition\footnote
{The main cooling
processes for the intrahalo gas are collisional excitation and ionization, 
recombination, and bremsstrahlung. The former is the most efficient for kinetic 
temperatures $T_k\approx 10^4-10^5$K and for neutral hydrogen and single ionized 
helium; for a meta--enriched gas, cooling is efficient at temperatures between 
$10^5-10^7$K. At higher temperatures, where the gas is completely ionized,
the dominant cooling process is bremsstrahlung. At temperatures lower than 
$10^4$K (small halos) and in absence of metals, the main cooling process is by 
$H_2$ and $HD$ molecule line emission.}. 
Since the seminal work by White \& Frenk (1990)
\cite{WF91}, the rate infall of gas available to form the galaxy is assumed
to be driven either by the free--fall time, $t_{ff}$, if $t_{ff}>t_{cool}$ 
or by the cooling time $t_{cool}$ if $t_{ff}<t_{cool}$.
The former case applies to halos of masses smaller than 
approximately $5\times 10^{11}\msun$, whilst the latter applies to more massive halos.
The cooling flow from the quasistatic hot atmosphere {\it is the process that
basically limits the baryonic mass of galaxies} \cite{silk77},
and therefore the bright end of the galaxy luminosity function; for the 
outer, dilute hot gas in large halos, $t_{cool}$ becomes larger than the Hubble
time. However, detailed calculations show that even so, in massive halos too much 
gas cools, and the bright end of the predicted luminosity function results
with a decrease slower than the observed one \cite{benson03}. Below we will see 
some solutions proposed to this problem.

More recently it was shown that the cooling of gas trapped in filaments
during the halo collapse may be so rapid that the gas flows along the 
filaments to the center, thus avoiding shock heating \cite{keres05}.
However, this process is efficient only for halos less massive than 
$2.5\times 10^{11}\msun$, which in any case (even if shock--heating happens),
cool their gas very rapidly \cite{bower05}. Thus, for modeling the formation
of disks, and for masses smaller than $\sim 5\times 10^{11}\msun$, we may assume 
that gas infalls in a dynamical time since the halo has virialized, or in 
two dynamical times since the protostructure was at its maximum expansion. 

\subparagraph{3. Disk formation, the origin of exponentially, and rotation curves.} 
The gas, originally distributed in mass and AM as the DM, cools and collapses
until it reaches centrifugal balance in a disk. Therefore, assuming detailed AM 
conservation, the radial mass distribution of the disk can be calculated by 
equating its specific AM to the AM of its final circular orbit in centrifugal 
equilibrium. The typical collapse factor of the gas within a DM halo is 
$\sim 10-15$\footnote{It is interesting to note that in 
the absence of a massive halo around galaxies, the collapse factor would 
be larger by $\sim M/M_d\approx 20$, where $M$ and $M_d$ are the total halo and 
disk masses, respectively \cite{padma}.}, depending on the initial halo
spin parameter $\lambda$; the higher the $\lambda$, the more extended (lower
surface density) is the resulting disk. The surface density profile of the disks 
formed within CDM halos is nearly exponential, which provides an explanation
to the long--standing question of why galaxy disks are exponential. This 
is a direct consequence of the AM distribution acquired by the halos
by tidal torques and mergers. In more
detail, however, the profiles are more concentrated in the center and with
a slight excess in the periphery than the exponential law \cite{FA00,Bullock01}. 
The cusp in the central disk could give rise to either a photometrical bulge
\cite{vdB01} or to a real kinematical bulge due to disk gravitational
instability enhanced by the higher central surface density \cite{AF00} (bulge 
secular formation). In a few cases (high--$\lambda$, low--concentrated halos),
purely exponential disks can be formed.
  
Baryons are a small mass fraction in the CDM halos, however, the disk formed
in the center is very dense (recall the high collapse factors), so that the 
contribution of the baryonic disk to the inner gravitational potential is 
important or even dominant. The formed disk will drag gravitationally DM,
producing an inner halo contraction that is important to calculate for 
obtaining the rotation curve decomposition. The method commonly used to
calculate it is based on the approximation of radial 
adiabatic invariance, where spherical symmetry and circular orbits
are assumed (e.g., \cite{flores,mo98}). However, the orbits in CDM
halos obtained in N--body simulations are elliptical rather than 
circular; by generalizing the adiabatic invariance to elliptical orbits,
the halo contraction becomes less efficient \cite{zavala,gnedin04}. 

The rotation curve decomposition of disks within contracted \Lcdm\ halos
are in general consistent with observations \cite{mo98,FA00,zavala}
(nearly--flat total rotation curves; maximum
disk for high--surface brightness disks; submaximum disk for the LSB
disks; in more detail, the outer rotation curve shape depends on
surface density, going from decreasing to increasing at the disk radius
for higher to lower densities, respectively). However, there are 
important non--solved issues. For example, from a large sample of
observed rotation curves, Persic et al. \cite{persic96} inferred
that the rotation curve shapes are described by an ``universal'' profile
that (i) depends on the galaxy luminosity and (ii) implies a halo 
profile different from the CDM (NFW) profile. Other studies confirm
only partially these claims \cite{verheijen97,zavala,catinella05}. 
Statistical studies of rotation curves are
very important for testing the \Lcdm\ scenario.

In general, the structure and dynamics of disks formed within \Lcdm\ halos
under the assumption of detailed AM conservation seem to be consistent with 
observations. An important result to remark is the successful prediction of the 
infrared Tully--Fisher relation and its scatter\footnote{In \S 4.1 we have 
shown that the basis of the Tully--Fisher relation is the CDM halo $M-V_m$ relation. 
From the pure halo to the disk+halo system there are several intermediate
processes that could distort the original $M-V_m$ relation. However, it
was shown that the way in which the CDM halo couples with the disk and the way
galaxies transform their gas into stars ``conspire'' to keep the relation.
Due to this conspiring, the Tully--Fisher relation is robust to variations in 
the baryon fraction $f_B$ (or mass--to--luminosity ratios) and in the spin 
parameter $\lambda$ \cite{FA00}.}. The core problem mentioned in \S 4.2 is 
the most serious potential difficulty.
Other potential difficulties are: (i) the predicted disk size 
(surface brightness) distribution implies a $P(\lambda)$ distribution narrower 
than that corresponding to \Lcdm\ halos by almost a factor of two 
\cite{laceydejong}; (ii) the internal AM distribution inferred from observations 
of dwarf galaxies seems not to be in agreement with the \Lcdm\ halo AM distribution 
\cite{vdBJ};  (iii) the inference of the halo profile from
the statistical study of rotation curve shapes seems not to be agreement with
CMD halos.
In N--body+hydrodynamical simulations of disk galaxy formation there was common
another difficulty called the 'angular momentum catastrophe': the simulated
disks ended too much concentrated, apparently due to AM transference of
baryons to DM during the gas collapse. The formation of highly concentrated
disks also affects the shape of the rotation curve (strongly decreasing), 
as well as the zero--point of the Tully--Fisher relation. Recent numerical 
simulations are showing that the 'angular momentum catastrophe',
rather than a physical problem, is a problem related to the resolution
of the simulations and the correct inclusion of feedback effects.

\subparagraph{4. Star formation and feedback.} 
We are coming to the less understood and most complicated aspects of the 
models of galaxy evolution, which deserve separate notes. The star
formation (SF) process is studied at two levels (each one by two separated
communities!): (i) the small--scale physics, related to the complex 
processes by which the cold gas inside molecular clouds fragments
and collapses into stars, and (ii) 
the large--scale physics, related
to the disk global instabilities that give rise to the largest
unities of SF, the molecular clouds. The SF physics incorporated to 
galaxy evolution models is still oversimplified, phenomenological
and refers to the latter item.
The large-scale SF cycle in normal galaxies is  believed to be 
self--regulated by a balance between the energy injection due to SF 
(mainly SNe) and dissipation (radiative or turbulent). Two main approaches 
have been used to describe the SF self--regulation in models of galaxy 
evolution: 
{\bf (a)} the halo cooling-feedback approach \cite{WF91}), 
{\bf (b)} the disk turbulent ISM approach \cite{FT94,WSilk94}.

According to the former, the cool gas is reheated by the ``galaxy'' SF 
feedback and driven back to the {\it intrahalo medium} until it again cools 
radiatively and collapses into the galaxy. This approach has been used in
semi--analytical models of galaxy formation where the internal structure
and hydrodynamics of the disks are not treated in detail. The
reheating rate is assumed to depend on the halo circular velocity $V_c$: 
$\dot{M}_{rh}\propto \dot{M}_s/V_c^{\alpha}$, where $\dot{M}_s$ is the 
SF rate (SFR) and $\alpha\ge 2$. Thus, the galaxy SFR, gas fraction 
and luminosity depend on $V_c$. In these models, the disk ISM is 
virtually ignored and the SN--energy injection is assumed to be as 
efficient as to reheat the cold gas up to the virial temperature
of the halo. A drawback of the model is that it predicts hot X-ray halos 
around disk galaxies much more luminous than those observed.

Approach (b) is more appropriate for models where the internal processes of 
the disk are considered. In this approach, the SF at a given radius $r$ 
is assumed to be triggered by disk gravitational instabilities (Toomre criterion)
and self--regulated by a balance between
energy injection (mainly by SNe) and dissipation in the turbulent ISM 
in the direction perpendicular to the disk plane:
\begin{eqnarray}
Q_g(r)\equiv \frac{v_g(r)\kappa(r)}{\pi G\Sigma_g(r)} < Q_{crit} \\
\gamma_{SN}\epsilon_{SN}\dot{\Sigma}_*(r) + \dot{\Sigma}_{E,accr}(r) =  
\frac{\Sigma_g(r) v_g^2(r)}{2t_{d}(r)}, 
\end{eqnarray}
where $v_g$ and $\Sigma_g$ are the gas velocity dispersion and surface density, 
$\kappa$ is the epicyclic frequency, $Q_{crit}$ is a critical value for
instability, $\gamma_{SN}$ and $\epsilon_{SN}$ are the kinetic
energy injection efficiency of the SN into the gas and the SN energy generated 
per gram of gas transformed into stars, respectively, $\dot{\Sigma}_*$
is the surface SFR, and $\dot{\Sigma}_{E,accr}$ is the kinetic energy input due to
mass accretion rate (or eventually any other energy source as AGN feedback). 
The key parameter in the self--regulating process is the dissipation time 
$t_d$. The disk ISM is a turbulent, non-isothermal, multi-temperature 
flow. Turbulent dissipation in the ISM is typically efficient 
($t_d\sim 10^7-10^8$yr) in such a way that
self--regulation happens at the characteristic vertical scales of the disk.
Thus, there is not too much room for strong feedback with the 
gas at heights larger than the vertical scaleheigth of normal present--day 
disks: self--regulation is at
the level of the disk, but not at the level of the gas corona around.
With this approach the predicted SFR is proportional to $\Sigma_g^n$
(Schmidt law), with $n\approx 1.4-2$ varying along the disk, in good
agreement with observational inferences. The typical 
SF timescales are not longer than $3-4$Gyr. Therefore, to keep
active SFRs in the disks, gas infall is necessary, a condition perfectly
fulfilled in the \Lcdm\ scenario.

Given the SFR radius by radius and time by time, and assuming an IMF,
the corresponding luminosities in different color bands can be calculated with 
stellar population synthesis models. The final result is then an evolving 
inside--out luminous disk with defined global and local colors. 

\paragraph{5. Secular evolution} 
The ``quiet'' evolution of galaxy disks as described above can be
disturbed by minor mergers (satellite accretion) and interactions 
with close galaxy companions. However, as several studies have shown, 
the disk may suffer even intrinsic instabilities which lead to secular 
changes in its structure, dynamics, and SFR. The main effects of
secular evolution, i.e. dynamical processes that act in a timescale 
longer than the disk dynamical time, are the vertical thickening
and ``heating'' of the disk, the formation of bars, which are efficient 
mechanisms of radial AM and mass redistribution, and the possible formation 
of (pseudo)bulges (see for recent reviews \cite{KK04,combes05}). Models of 
disk galaxy evolution should include these processes, which also can 
affect disk properties, for example increasing the disk scale radii \cite{val03}.

\subsection{Spheroids}

As mentioned in \S 2, the simple appearance, the dominant old stellar populations,
the $\alpha$--elements enhancement, and the dynamically hot structure of 
spheroids suggest that they were formed by an early ($z\grtsim 4$) single violent event 
with a strong burst of star formation, followed by passive evolution of their 
stellar population ({\it monolithic} mechanism). Nevertheless, both observations 
and theory point out to a more complex situation.  There are two ways to define 
the formation epoch of a spheroid: when most of its stars formed or when the stellar 
spheroid acquired its dynamical properties in violent or secular processes.
For the monolithic collapse mechanism both epochs coincide. 

In the context of the \Lcdm\ scenario, spheroids are expected to be formed 
basically as the result of major mergers of disks. However, 

{\narrower
$\bullet$ if the major mergers occur at high redshifts, when the disks are 
mostly gaseous, then the situation is close to the monolithic collapse;

$\bullet$ if the major mergers occur at low redshifts, when the galaxies 
have already transformed a large fraction of their gas into stars, then the spheroids
assemble by the ``classical'' dissipationless collision.
\par }

Besides, stellar disks may develop spheroids in their centers (bulges) by
secular evolution mechanisms, both intrinsic or enhanced by minor mergers and
interactions; this channel of spheroid formation should work for 
late--type galaxies and it is supported by a large body of observations 
\cite{KK04}. But the picture is even more complex in the hierarchical 
cosmogony as galaxy morphology may be continuously changing, depending on the 
MAH (smooth accretion and violent mergers) and environment. An spheroid
formed early should continue accreting gas so that a new, younger disk
grows around. A naive expectation in the context of the \Lcdm\ scenario
is that massive elliptical galaxies should be assembled mainly by late major 
mergers of the smaller galaxies in the hierarchy. It is also expected that 
the disks in galaxies with small bulge--to--disk
ratios should be on average redder than those in galaxies with large
bulge--to--disk ratios, contrary to observations.  

Although it is currently subject of debate, a more elaborate
picture of spheroid formation is emerging now in the context of the \Lcdm\
hierarchical scenario (see \cite{silkrees,FA03,delucia05} and the references therein). 
The basic ideas are that massive ellipticals formed early ($z\grtsim 3$) 
and in a short timescale by the merging of gas--rich disks in rare high--peak, 
clustered regions of the 
Universe. The complex physics of the merging implies (i) an ultraluminous 
burst of SF obscured by dust (cool ULIRG phase) and the establishment of a spheroidal
structure, (ii) gas collapse to the center, a situation that favors the 
growth of the preexisting massive black hole(s) through an Eddington or even
super--Eddington regime (warm ULIRG phase), (iii) the switch on of the AGN 
activity associated to the supermassive black hole when reaching a critical 
mass, reverting then the gas inflow to gas outflow (QSO phase), (iv) the switch 
off of the AGN activity leaving a giant stellar spheroid with a supermassive 
black hole in the center and a hot gas corona around (passive elliptical 
evolution). In principle, the hot corona may cool by cooling flows and
increase the mass of the galaxy, likely renewing a disk around the spheroid.
However, it seems that recurrent AGN phases (less energetic than the initial
QSO phase) are possible during the life of the spheroid. Therefore, the 
energy injected from AGN in the form of radio jets (feedback) can be responsible
for avoiding the cooling flow. This way is solved the problem of disk
formation around the elliptical, as well as the problem of the extended bright
end in the luminosity function. It is also important to note that
as soon as the halo hosting the elliptical becomes a subhalo of the group
or cluster, the MAH is truncated (\S 4). According to the model just
described, massive elliptical galaxies were in place at high redshifts,
while less massive galaxies (collapsing from more common density peaks)
assembled later. This model was called {\it downsizing} or anti-hierarchical.
In spite of the name, it fits perfectly within the hierarchical \Lcdm\ scenario.

\subsection{Drivers of the Hubble sequence}

$\bullet$ Disks are generic objects formed by gas dissipation and collapse
inside the growing CDM halos. Three (cosmological) initial and
boundary conditions related to the halos define the main
properties of disks in isolated halos:

{\narrower
1. The virial mass, which determines extensive properties

2. The spin parameter $\lambda$, which determines mainly the 
disk surface brightness (SB; it gives rise to the sequence from
high SB to low SB disks) and strongly influences the rotation
curve shape and the bulge--to--disk ratio (within the secular scenario).
$\lambda$ also plays some role in the SFR history. 

3. The MAH, which drives the gas infall rate and, therefore,
the disk SFR and color; the MAH determines also the halo
concentration, and its scatter is reflected in the scatter
of the Tully--Fisher relation.

\par }

The two latter determine the intensive properties of disks,
suggesting a biparametrical sequence in SB and color. 
There is a fourth important parameter, the galaxy baryon
fraction $f_B$, which influences the disk SB and rotation
curve shape. We have seen that $f_B$ in galaxies is
3--5 times lower than the universal $\Omega_B/\Omega_{DM}$
fraction. This parameter is related probably to astrophysical
processes as gas dissipation and feedback.  

$\bullet$ The clustering of CDM halos follows an spatial distribution
with very different large--scale environments. In low--density
environments, halos live mostly isolated, favoring the formation
of disks, whose properties are driven by the factors mentioned
above. However, as we move to higher--density environments, 
halos form from more and more clustered high--peak perturbations 
that assemble early by violent major mergers: this is the necessary
condition to form massive ellipticals. At some time, the larger scale in 
the hierarchy collapses and the halo becomes a subhalo: the mass aggregation 
is then truncated and the probability of merging decreases dramatically. 
Elliptical galaxies are settled and continue evolving passively. 
Thus, the environment of CDM halos is another important driver of the
Hubble sequence, able to establish the main body of the observed blue--red
and early--type morphology sequences and their dependences on density. 

$\bullet$ Although the initial, boundary and environmental conditions
provided by the \Lcdm\ scenario are drivers of several of the main
properties and correlations of galaxies, astrophysical processes should also
play an important role. The driving astrophysical processes are
global SF and feedback. They should come in two modes that drive
the disk and elliptical sequences: (i) the quiescent
disk mode, where disk instabilities trigger SF and local (negative) feedback 
self--regulates the SFR, and (ii) the bursting mode of violent mergers
of gaseous galaxies, where local shocks and gravothermal catastrophe trigger
SF, and presumably a positive feedback increases its efficiency. 
Other important astrophysical drivers of galaxy properties are: 
(i)  the SN--induced wind--driven outflows, which are important to shape
the properties of dwarf galaxies ($M\lesssim 10^{10}\msun$, $V_m\lesssim 80$km/s),
(ii) the AGN--induced hydrodynamical outflows, which are important
to prevent cooling flows in massive ellipticals, (iii) several processes
typical of high--density environments such as ram pressure, harassment, strangulation,
etc., presumably important to shape some properties of galaxies in clusters.

\section{Issues and outlook}

Our understanding of galaxy formation and evolution is in its infancy.
So far, only the first steps were given in the direction of consolidating
a theory in this field. The process is apparently so complex and non--linear 
that several specialists do not expect the emergence of a theory in the 
sense that a few driving parameters and factors might explain
the main body of observations. Instead, the most popular trend now is to
attain some description of galaxy evolution by simulating it in
expensive computational runs. I believe that simulations are a 
valuable tool to extend a bridge between reality and the 
distorted (biased) information given by observations. However,
the search of basic theories for explaining galaxy formation
and evolution should not be replaced by the only effort of simulating
in detail what in fact we want to get. The power of science lies
in its predictive capability. Besides, if galaxy theory becomes predictive,
then its potential to test fundamental and cosmological theories
will be enormous.  

Along this notes, potential difficulties or unsolved problems
of the \Lcdm\ scenario were discussed. Now I summarize and 
complement them:

\paragraph{Physics} 

$\bullet$ What is non--baryonic DM? From the structure formation 
side, the preferred (and necessary!) type is CDM, though WDM with filtering
masses below $\sim 10^9\msun$ is also acceptable. So far none
of the well--motivated cold or warm non--baryonic particles 
have been detected in Earth experiments. The situation is even 
worth for proposals not based on elemental particles as 
DM from extra--dimensions.   
 
$\bullet$ What is Dark Energy? Dark Energy does not play apparently
a significant role in the internal evolution of perturbations but
it crucially defines the cosmic timescale and expansion rate, which  
are important for the growing factor of perturbations. 
The simplest interpretation of Dark Energy is the homogeneous and inert
cosmological constant $\Lambda$, with equation of state parameter $w=-1$ and
$\rho_{\Lambda}=$const.  The combinations of different
cosmological probes tend to favor the flat-geometry $\Lambda$ models
with ($\Omega_M, \Omega_\Lambda$)$\approx$(0.26, 0.74). However, the 
cosmological constant explanation of Dark Energy faces serious theoretical 
problems. Several alternatives to $\Lambda$ were proposed to
ameliorate partially these problems (e.g. quintaessence,
k--essence, Chaplygin gas, etc.).  Also have been proposed unifying schemes
of DM and Dark Energy through scalar fields (e.g, \cite{matos}). 

\paragraph{Cosmology}

$\bullet$ Inflation provides a natural mechanism for the generation
of primordial fluctuations. The nearly scale--invariance of the 
primordial power spectrum is well predicted by several inflation models, 
but its amplitude, rather than being predicted, is empirically inferred
from observations of CMBR anisotropies. Another aspect of 
primordial fluctuations not well understood is related to their
statistics, i.e.,  whether they are Gaussian--distributed or not.
And this is crucial for cosmic structure formation.

$\bullet$ Indirect pieces of evidence are consistent with the main 
predictions of inflation regarding primordial fluctuations. However, 
more direct tests of this theory are highly desirable. Hopefully, 
CMBR anisotropy observations will allow for some more direct tests
(e.g., effects from primordial gravitational waves).

\paragraph{Astrophysics}

$\bullet$  Issues at small scales. The excess of substructure
(satellite galaxies) can be apparently solved by inhibition of galaxy formation
in small halos due to UV--radiation produced by reionization and due to feedback,
rather than to modifications to the scenario (e.g., the introduction of WDM).
Observational inferences of the inner volume and phase--space densities of 
dwarf satellite galaxies are crucial to explore this question. The direct
detection (with gravitational lensing) of the numerous subhalo (dark galaxy) 
population predicted by CDM for the Galaxy halo is a decisive test on the 
problem of substructure.  The CDM prediction of cuspy halos is a more 
involved problem when confronting it with observational inferences. If the disagreement
persists, then either the $\Lambda$CDM scenario will need a modification
(e.g., introduction of self--interaction or annihilation), or astrophysical processes
involving gas baryon physics should be in action. However, there are still
unsolved issues at the intermediate level: for example, the central halo density 
profile of galaxies is inferred from observations of inner rotation curves 
{\it under} several assumptions that could be incorrect. An interesting technique 
to overcome this problem is being currently developed: to simulate as realistically as 
possible a given galaxy, ``observe'' its rotation curve and then compare with that 
of the real galaxy (see \S\S 4.1).    

$\bullet$ The early formation of massive red elliptical galaxies can be accommodated
in the hierarchical \Lcdm\ scenario (\S\S 5.2) {\it if} spheroids are produced by
the major merger of gaseous disks, and if the cold gas is transformed rapidly 
into stars during the merger in a dynamical time or so. Both conditions should
be demonstrated, in particular the latter. A kind of positive feedback
seems to be necessary for such an efficient star formation rate  (ISM shocks 
produced by the jets generated in the vicinity of supermassive black holes?). 

$\bullet$ Once the elliptical has formed early, the next difficulty is how to avoid
further (disk) growth around it. The problem can be partially solved by considering
that ellipticals form typically in dense, clustered environments,
and at some time they become substructures of larger virialized groups or clusters,
truncating any possible accretion to the halo/galaxy. However, (i) galaxy halos,
even in clusters, are filled with a reservoir of gas, and (ii) there are some
ellipticals in the field. Therefore, negative feedback mechanisms are needed
to stop gas cooling and accretion. AGN--triggered radio jets have been proposed as 
a possible mechanism, but further investigation is necessary. 

$\bullet$ The merging mechanism of bulge formation within the hierarchical model
implies roughly bluer (later formed) disks as the bulge--to--disk ratio is larger,
contrary to the observed trend. The secular scenario could solve this problem
but it is not still clear whether bars disolve or not in favor of pseudobulges.
It is not clear also if the secular scenario could predict the central supermassive
black hole mass--velocity dispersion relation.  

$\bullet$ We lack a fundamental theory of star formation. So far, simple
models, or even just phenomenological recipes, have been used in galaxy formation 
studies. The two proposed modes of star formation (the quiescent, inefficient, 
disk self--regulated regime, and the violent efficient star--bursting regime 
in mergers) are oversimplifications of a much more complex problem with
more physical mechanisms (shocks, turbulence, etc.). Closely related to star 
formation is the problem of feedback. The feedback mechanisms are different
in the ISM of disks, in the gaseous medium of merging galaxies with a powerful 
energy source (the AGN) other than stars, and in the diluted and hot intrahalo 
medium around galaxies.   
 
$\bullet$ We have seen in \S\S 2.2 that at the present epoch only $\approx 9\%$
of baryons are within virialized structures. Where are the remaining $91\%$ of the 
baryons? The fraction of particles
in halos measured in \Lcdm\ N--body cosmological simulations is $\sim 50\%$. 
This sounds good but still we have to explain, within the \Lcdm\ scenario,
the $\sim 40\%$ of missing baryons. The question is were these baryons
never trapped by collapsed halos or were they trapped but later
expelled due to galaxy feedback. Large--scale N--body+hdydrodynamical 
simulations have shown that the gravitational collapse of filaments may heat
the gas and keep a big fraction of baryons outside the collapsed halos \cite{dave}.
Nevertheless, feedback mechanisms, especially at high redshifts, are also
predicted to be strong enough as to expel enriched gas back to the
Intergalactic Medium. The problem is open.

The field has plenty of open and exciting problems. The \Lcdm\ scenario
has survived many observational tests but it still faces the difficulties
typical of a theory constructed phenomenologically and heuristically.
Even if in the future it is demonstrated that CDM does not exist (which
is little probable), the \Lcdm\ scenario would serve as an excellent ``fitting''
model to reality, which would strongly help researchers in developing new 
theories. 

\subparagraph{Acknowledgments.-} I am in debt with Dr. I. Alc\'antara-Ayala and
R. N\'u\~nez-L\'opez for their help in the preparation of the figures. I am also 
grateful to J. Benda for grammar corrections, and to the Editors for their 
infinite patience. 

%


\end{document}